\documentclass[aps,nofootinbib,preprint,superscriptaddress]{revtex4}%
\usepackage{hyperref}
\usepackage{amsmath}
\usepackage{amsfonts}
\usepackage{amssymb}
\usepackage{graphicx,subfig}
\usepackage{color}
\usepackage{graphicx}%
\providecommand{\U}[1]{\protect\rule{.1in}{.1in}}

\begin{document}
\title{Recent developments of the Lauricella string scattering amplitudes and their
exact $SL(K+3,C)$ Symmetry}
\author{Sheng-Hong Lai}
\email{xgcj944137@gmail.com}
\affiliation{Department of Electrophysics, National Chiao-Tung University, Hsinchu, Taiwan, R.O.C.}
\author{Jen-Chi Lee}
\email{jcclee@cc.nctu.edu.tw}
\affiliation{Department of Electrophysics, National Chiao-Tung University, Hsinchu, Taiwan, R.O.C.}
\author{Yi Yang}
\email{yiyang@mail.nctu.edu.tw}
\affiliation{Department of Electrophysics, National Chiao-Tung University, Hsinchu, Taiwan, R.O.C.}
\author{}
\date{\today }

\begin{abstract}
In this review we propose a new perspective to demonstrate Gross conjecture on
high energy symmetry of string theory \cite{GM,GM1,Gross,Gross1,GrossManes}.
We review the construction of the exact string scattering amplitudes (SSA) of
three tachyons and one arbitrary string state, or the Lauricella SSA (LSSA),
in the $26D$ open bosonic string theory. These LSSA form an infinite
dimensional representation of the $SL(K+3,%
\mathbb{C}
)$ group. Moreover, we show that the $SL(K+3,%
\mathbb{C}
)$ group can be used to solve all the LSSA and express them in terms of one
amplitude. As an application in the hard scattering limit, the LSSA can be
used to directly prove Gross conjecture which was previously corrected and
proved by the method of decoupling of zero norm states (ZNS)
\cite{Lee,LeePRL,lee-Ov,ChungLee1,ChanLee1,ChanLee,ChanLee2,CHL,CHLTY2,
CHLTY1,susy}. Finally, the exact LSSA can be used to rederive the recurrence
relations of SSA in the Regge scattering limit with associated $SL(5,%
\mathbb{C}
)$ symmetry and the extended recurrence relations (including the mass and spin
dependent string BCJ relations) in the nonrelativistic scattering limit with
associated $SL(4,%
\mathbb{C}
)$ symmetry discovered recently.

\end{abstract}
\maketitle
\tableofcontents

%

\setcounter{equation}{0}
\renewcommand{\theequation}{\arabic{section}.\arabic{equation}}%

\section{Introduction}

In contrast to low-energy string theory, many issues regarding high-energy behavior
of string theory have not yet been well understood. Historically, it
was first conjectured by Gross \cite{GM,GM1,Gross,Gross1,GrossManes} that
there exist infinite linear relations among hard string scattering amplitudes
(HSSA) of different string states. Moreover, these linear relations are so
powerful that they can be used to solve all HSSAs and express them in terms of
one amplitude. This conjecture was later (slightly) corrected and proved by
using the decoupling of zero norm states \cite{Lee,LeePRL,lee-Ov,ChungLee1} in
\cite{ChanLee1,ChanLee,ChanLee2,CHL,CHLTY2, CHLTY1,susy}. For more details,
see the recent review articles \cite{review, over}.

In this paper, we review another perspective to understand the high-energy
behavior of strings and demonstrate the Gross conjecture regarding the high-energy symmetry of
string theory. Since the theory of strings, as a quantum theory, consists of
an infinite number of particles with arbitrarily high spins and masses, one first
crucial step to uncovering its high-energy behavior is to exactly calculate a
class of SSA that contains the whole spectrum valid for all energies. Recently,
the present authors constructed a class of such an exact SSA that contains three
tachyons and one arbitrary string state in the spectrum, or the Lauricella SSA
(LSSA), in the $26D$ open bosonic string theory.

There are many works based on the research of tensionless strings ($\alpha^{\prime
}\rightarrow\infty$)
\cite{less1,less2,less3,less5,less6,less7,less8,less9,less10,less11,less12}
that are related to our works on high-energy symmetry of string theory.
However, as presented in Section 4, in our high-energy calculation, we
keep the mass level parameter $M$ of the string spectrum fixed as a finite
constant at each mass level. In contrast, in the calculation of
tensionless strings in the literature, all string states are massless in the
limit $\alpha^{\prime}\rightarrow\infty$. We believe that by keeping $M$ fixed
as a finite constant, one can obtain more information about the high-energy
behavior of string theory.

More recently, other interesting approaches have been proposed in the literature which
deal with higher spin string states \cite{spin1,spin2,spin3,spin4,spin5,spin6}. More works need to be done on higher spin string states, especially higher
massive fermionic string states in the R-sector of superstrings, before one can
fully understand the high-energy behavior of superstring theory.

In Section 2 of this review, we calculate the LSSAs and express them in terms
of $D$-type Lauricella functions. As an application, we easily reproduce
the string BCJ 
relation \cite{Closed,LLY,stringBCJ, stringBCJ2}. As
an illustration of LSSA, we give two simple examples to demonstrate the
complicated notation. We then proceed to show that the LSSAs form an infinite
dimensional representation of the $SL(K+3,C)$ group. For simplicity, and as an
warm up exercise, we begin with the case of $K=1$ or the $SL(4,C)$ group.

In Section 3, we first show that there exist $K+2$ recurrence relations
among the $D$-type Lauricella functions. We then show that the corresponding
$K+2$ recurrence relations among the LSSAs can be used to reproduce the Cartan
subalgebra and simple root system of the $SL(K+3,%
\mathbb{C}
)$ group with rank $K+2$. As a result, the $SL(K+3,%
\mathbb{C}
)$ group can be used to solve all the LSSAs and express them in terms of one
amplitude. We stress that these exact nonlinear relations among the
exact LSSAs are generalizations of the linear relations among HSSAs in the hard
scattering limit conjectured by Gross. Finally, we show that, for the first few
mass levels, the Lauricella recurrence relations imply the validity of Ward
identities derived from the decoupling of Lauricella ZNS. 
However, these
Lauricella Ward identities are not good enough to solve all the LSSAs
and express them in terms of one amplitude.

In Section 4 of this review, we calculate symmetries or relations among the
LSSAs of different string states at various scattering limits. These include
the linear relations first conjectured by Gross
\cite{GM,GM1,Gross,Gross1,GrossManes} and later corrected and proved in
\cite{ChanLee1,ChanLee2,CHL,CHLTY2, CHLTY1,susy} in the hard scattering limit,
the recurrence relations in the Regge scattering limit with associated $SL(5,%
\mathbb{C}
)$ symmetry \cite{KLY,LY,AppellLY} and the extended recurrence relations
(including the mass and spin dependent string BCJ relations) in the
nonrelativistic scattering limit with associated $SL(4,%
\mathbb{C}
)$ symmetry \cite{LLY} discovered recently.

In Section 5, we give a brief conclusion and suggest some future works.
Finally, in the appendix, we present detailed calculations of the LSSAs presented in Section 2 of the text.%


\section{The Exact LSSAs and Their \boldmath{$SL(K+3,C)$} Symmetry}

\subsection{The Exact LSSAs}

One important observation of calculating LSSAs is to first note that the SSAs of
three tachyons and one arbitrary string state with polarizations orthogonal to
the scattering plane vanish. This observation greatly simplifies the
calculation of the LSSA. In the CM 
frame, we define the kinematics {as}
\begin{align}
k_{1}  &  =\left(  \sqrt{M_{1}^{2}+|\vec{k_{1}}|^{2}},-|\vec{k_{1}}|,0\right)
,\\
k_{2}  &  =\left(  \sqrt{M_{2}+|\vec{k_{1}}|^{2}},+|\vec{k_{1}}|,0\right)  ,\\
k_{3}  &  =\left(  -\sqrt{M_{3}^{2}+|\vec{k_{3}}|^{2}},-|\vec{k_{3}}|\cos
\phi,-|\vec{k_{3}}|\sin\phi\right)  ,\\
k_{4}  &  =\left(  -\sqrt{M_{4}^{2}+|\vec{k_{3}}|^{2}},+|\vec{k_{3}}|\cos
\phi,+|\vec{k_{3}}|\sin\phi\right)  \label{CMframe}%
\end{align}
with $M_{1}^{2}=M_{3}^{2}=M_{4}^{2}=-2$ and $\phi$ is the scattering angle.
The Mandelstam variables are $s=-\left(  k_{1}+k_{2}\right)  ^{2}$,
$t=-\left(  k_{2}+k_{3}\right)  ^{2}$ and $u=-\left(  k_{1}+k_{3}\right)
^{2}$. There are three polarizations on the scattering plane, and they are
defined to be \cite{ChanLee1,ChanLee2}%
\vspace{6pt}
\begin{align}
e^{T}  &  =(0,0,1),\label{aa}\\
e^{L}  &  =\frac{1}{M_{2}}\left(  |\vec{k_{1}}|,\sqrt{M_{2}+|\vec{k_{1}}|^{2}%
},0\right)  ,\label{bb}\\
e^{P}  &  =\frac{1}{M_{2}}\left(  \sqrt{M_{2}+|\vec{k_{1}}|^{2}},|\vec{k_{1}%
}|,0\right)  \label{cc}%
\end{align}
where $e^{P}=\frac{1}{M_{2}}(E_{2},\mathrm{k}_{2},0)=\frac{k_{2}}{M_{2}}$ is the
momentum polarization, $e^{L}=\frac{1}{M_{2}}(\mathrm{k}_{2},E_{2},0)$ is the
longitudinal polarization and $e^{T}=(0,0,1)$ is the transverse polarization. For
later use, we also define%
\begin{equation}
k_{i}^{X}\equiv e^{X}\cdot k_{i}\text{ \ for \ }X=\left(  T,P,L\right)  .
\end{equation}

We now proceed to calculate the LSSAs of three tachyons and one
arbitrary string state in the $26D$ open bosonic string theory. The general
states at mass level $M_{2}^{2}=2(N-1)$, $N=\sum_{n,m,l>0}\left(  nr_{n}%
^{T}+mr_{m}^{P}+lr_{l}^{L}\right)  $ with polarizations on the scattering
plane are of the following form:%
\begin{equation}
\left\vert r_{n}^{T},r_{m}^{P},r_{l}^{L}\right\rangle =\prod_{n>0}\left(
\alpha_{-n}^{T}\right)  ^{r_{n}^{T}}\prod_{m>0}\left(  \alpha_{-m}^{P}\right)
^{r_{m}^{P}}\prod_{l>0}\left(  \alpha_{-l}^{L}\right)  ^{r_{l}^{L}}%
|0,k\rangle. \label{state}%
\end{equation}

The $\left(  s,t\right)  $ channel of the LSSA can be calculated to be
\cite{LLY2}
\begin{align}
A_{st}^{(r_{n}^{T},r_{m}^{P},r_{l}^{L})}  &  =\prod_{n=1}\left[
-(n-1)!k_{3}^{T}\right]  ^{r_{n}^{T}}\cdot\prod_{m=1}\left[  -(m-1)!k_{3}%
^{P}\right]  ^{r_{m}^{P}}\prod_{l=1}\left[  -(l-1)!k_{3}^{L}\right]
^{r_{l}^{L}}\nonumber\\
&  \cdot B\left(  -\frac{t}{2}-1,-\frac{s}{2}-1\right)  F_{D}^{(K)}\left(
-\frac{t}{2}-1;R_{n}^{T},R_{m}^{P},R_{l}^{L};\frac{u}{2}+2-N;\tilde{Z}_{n}%
^{T},\tilde{Z}_{m}^{P},\tilde{Z}_{l}^{L}\right)  \label{st1}%
\end{align}
where we have defined%
\begin{equation}
R_{k}^{X}\equiv\left\{  -r_{1}^{X}\right\}  ^{1},\cdots,\left\{  -r_{k}%
^{X}\right\}  ^{k}\text{ \ with \ }\left\{  a\right\}  ^{n}%
=\underset{n}{\underbrace{a,a,\cdots,a}}.
\end{equation}
and%
\begin{equation}
Z_{k}^{X}\equiv\left[  z_{1}^{X}\right]  ,\cdots,\left[  z_{k}^{X}\right]
\text{ \ \ with \ \ }\left[  z_{k}^{X}\right]  =z_{k0}^{X},\cdots,z_{k\left(
k-1\right)  }^{X}. \label{comp}%
\end{equation}

In Equation (\ref{comp}), we have defined
\begin{align}
z_{k}^{X}  &  =\left\vert \left(  -\frac{k_{1}^{X}}{k_{3}^{X}}\right)
^{\frac{1}{k}}\right\vert ,\ z_{kk^{\prime}}^{X}=z_{k}^{X}e^{\frac{2\pi
ik^{\prime}}{k}},\ \tilde{z}_{kk^{\prime}}^{X}\equiv1-z_{kk^{\prime}}%
^{X}\text{ \ \ for \ \ }k^{\prime}=0,\cdots,k-1\\
\text{or \ }\left[  z_{k}^{X}\right]   &  =z_{k}^{X},z_{k}^{X}\omega
_{k},...,z_{k}^{X}\omega_{k}^{k-1},\text{ \ \ \ }\omega_{k}=e^{\frac{2\pi
i}{k}}.
\end{align}

The integer $K$ in Equation (\ref{st1}) is defined to be%
\begin{equation}
\text{ }K=\underset{\{\text{for all }r_{j}^{T}\neq0\}}{\sum j}%
+\underset{\{\text{for all }r_{j}^{P}\neq0\}}{\sum j}+\underset{\{\text{for
all }r_{j}^{L}\neq0\}}{\sum j}. \label{kk}%
\end{equation}

The $D$-type Lauricella function $F_{D}^{(K)}$ in Equation (\ref{st1}) is one of the
four extensions of the Gauss hypergeometric function to $K$ variables and is
defined to be%
\begin{align}
&  F_{D}^{(K)}\left(  \alpha;\beta_{1},...,\beta_{K};\gamma;x_{1}%
,...,x_{K}\right) \nonumber\\
&  =\sum_{n_{1},\cdots,n_{K}=0}^{\infty}\frac{\left(  \alpha\right)
_{n_{1}+\cdots+n_{K}}}{\left(  \gamma\right)  _{n_{1}+\cdots+n_{K}}}%
\frac{\left(  \beta_{1}\right)  _{n_{1}}\cdots\left(  \beta_{K}\right)
_{n_{K}}}{n_{1}!\cdots n_{K}!}x_{1}^{n_{1}}\cdots x_{K}^{n_{K}}%
\end{align}
where $(\alpha)_{n}=\alpha\cdot\left(  \alpha+1\right)  \cdots\left(
\alpha+n-1\right)  $ is the Pochhammer symbol. An integral
representation of the Lauricella function $F_{D}^{(K)}$ was discovered by Appell
and Kampe de Feriet (1926) \cite{Appell},%
\begin{align}
&  F_{D}^{(K)}\left(  \alpha;\beta_{1},...,\beta_{K};\gamma;x_{1}%
,...,x_{K}\right) \nonumber\\
&  =\frac{\Gamma(\gamma)}{\Gamma(\alpha)\Gamma(\gamma-\alpha)}\int_{0}%
^{1}dt\,t^{\alpha-1}(1-t)^{\gamma-\alpha-1}\cdot(1-x_{1}t)^{-\beta_{1}%
}(1-x_{2}t)^{-\beta_{2}}...(1-x_{K}t)^{-\beta_{K}},
\end{align}
which was used to calculate Equation (\ref{st1}).

\subsection{String BCJ Relation as a By-Product}

Alternatively, by using the identity of the Lauricella function for $b_{i}\in
Z^{-}$,%
\begin{align}
&  F_{D}^{(K)}\left(  a;b_{1},...,b_{K};c;x_{1},...,x_{K}\right)
=\frac{\Gamma\left(  c\right)  \Gamma\left(  c-a-\sum b_{i}\right)  }%
{\Gamma\left(  c-a\right)  \Gamma\left(  c-\sum b_{i}\right)  }\nonumber\\
\cdot &  F_{D}^{(K)}\left(  a;b_{1},...,b_{K};1+a+\sum b_{i}-c;1-x_{1}%
,...,1-x_{K}\right)  ,
\end{align}
one can rederive the string BCJ relations \cite{Closed,LLY,stringBCJ,
stringBCJ2}:
\begin{align}
\frac{A_{st}^{(r_{n}^{T},r_{m}^{P},r_{l}^{L})}}{A_{tu}^{(r_{n}^{T},r_{m}%
^{P},r_{l}^{L})}}  &  =\frac{(-)^{N}\Gamma\left(  -\frac{s}{2}-1\right)
\Gamma\left(  \frac{s}{2}+2\right)  }{\Gamma\left(  \frac{u}{2}+2-N\right)
\Gamma\left(  -\frac{u}{2}-1+N\right)  }\nonumber\\
&  =\frac{\sin\left(  \frac{\pi u}{2}\right)  }{\sin\left(  \frac{\pi s}%
{2}\right)  }=\frac{\sin\left(  \pi k_{2}\cdot k_{4}\right)  }{\sin\left(  \pi
k_{1}\cdot k_{2}\right)  }. \label{BCJ}%
\end{align}
This gives another form of the $\left(  s,t\right)  $ channel amplitude:
\begin{align}
&  A_{st}^{(r_{n}^{T},r_{m}^{P},r_{l}^{L})}\nonumber\\
&  =B\left(  -\frac{t}{2}-1,-\frac{s}{2}-1+N\right)  \prod_{n=1}\left[
-(n-1)!k_{3}^{T}\right]  ^{r_{n}^{T}}\nonumber\\
&  \cdot\prod_{m=1}\left[  -(m-1)!k_{3}^{P}\right]  ^{r_{m}^{P}}\prod
_{l=1}\left[  -(l-1)!k_{3}^{L}\right]  ^{r_{l}^{L}}\nonumber\\
&  \cdot F_{D}^{(K)}\left(  -\frac{t}{2}-1;R_{n}^{T},R_{m}^{P},R_{l}^{L}%
;\frac{s}{2}+2-N;Z_{n}^{T},Z_{m}^{P},Z_{l}^{L}\right)  . \label{st2}%
\end{align}

Similarly, the $\left(  t,u\right)  $ channel amplitude can be calculated to
be
\begin{align}
&  A_{tu}^{(r_{n}^{T},r_{m}^{P},r_{l}^{L})}\nonumber\\
&  =B\left(  -\frac{t}{2}-1,-\frac{u}{2}-1\right)  \prod_{n=1}\left[
-(n-1)!k_{3}^{T}\right]  ^{r_{n}^{T}}\nonumber\\
&  \cdot\prod_{m=1}\left[  -(m-1)!k_{3}^{P}\right]  ^{r_{m}^{P}}\prod
_{l=1}\left[  -(l-1)!k_{3}^{L}\right]  ^{r_{l}^{L}}\nonumber\\
&  \cdot F_{D}^{(K)}\left(  -\frac{t}{2}-1;R_{n}^{T},R_{m}^{P},R_{l}^{L}%
;\frac{s}{2}+2-N;Z_{n}^{T},Z_{m}^{P},Z_{l}^{L}\right)  . \label{tu2}%
\end{align}

The detailed calculation of all the above results can be found in the
appendix. To illustrate the complicated notations used in Equation (\ref{st1}), we
give two explicit examples of the LSSA in the following subsection.

\subsection{Two Simple Examples of the LSSA}

\subsubsection{Example One}

We take the tensor state of the second vertex to be%
\begin{equation}
\left\vert \text{state}\right\rangle =\left(  \alpha_{-1}^{T}\right)
^{r_{1}^{T}}\left(  \alpha_{-1}^{P}\right)  ^{r_{1}^{P}}\left(  \alpha
_{-1}^{L}\right)  ^{r_{1}^{L}}|0,k\rangle.
\end{equation}
The LSSA in Equation (\ref{st1}) can then be calculated to be%
\begin{align}
A_{st}^{(r_{1}^{T},r_{1}^{P},r_{l}^{L})}  &  =\left(  -k_{3}^{T}\right)
^{r_{1}^{T}}\left(  -k_{3}^{P}\right)  ^{r_{1}^{P}}\left(  -k_{3}^{L}\right)
^{r_{1}^{L}}B\left(  -\frac{t}{2}-1,-\frac{s}{2}-1\right) \nonumber\\
&  \cdot F_{D}^{(3)}\left(  -\frac{t}{2}-1;-r_{1}^{T},-r_{1}^{P},-r_{1}%
^{L};\frac{u}{2}+2-N;\tilde{z}_{10}^{T},\tilde{z}_{10}^{P},\tilde{z}_{10}%
^{L}\right)
\end{align}
where the arguments in $F_{D}^{(3)}$ are calculated to be%
\begin{align}
R_{n}^{T}  &  =\left\{  -r_{1}^{T}\right\}  ^{1},\cdots,\left\{  -r_{n}%
^{T}\right\}  ^{k}=\left\{  -r_{1}^{T}\right\}  ^{1}=-r_{1}^{T},\nonumber\\
R_{m}^{P}  &  =\left\{  -r_{1}^{P}\right\}  ^{1},\cdots,\left\{  -r_{m}%
^{P}\right\}  ^{k}=\left\{  -r_{1}^{P}\right\}  ^{1}=-r_{1}^{P},\nonumber\\
R_{l}^{L}  &  =\left\{  -r_{1}^{L}\right\}  ^{1},\cdots,\left\{  -r_{l}%
^{L}\right\}  ^{k}=\left\{  -r_{1}^{L}\right\}  ^{1}=-r_{1}^{L},\\
\tilde{Z}_{n}^{T}  &  =\left[  \tilde{z}_{1}^{T}\right]  ,\cdots,\left[
\tilde{z}_{n}^{T}\right]  =\left[  \tilde{z}_{1}^{T}\right]  =\tilde{z}%
_{10}^{T}=1-z_{10}^{T}=1-z_{k}^{T}e^{\frac{2\pi i0}{1}}=1-\left\vert
-\frac{k_{1}^{T}}{k_{3}^{T}}\right\vert ,\nonumber\\
\tilde{Z}_{n}^{P}  &  =\left[  \tilde{z}_{1}^{P}\right]  ,\cdots,\left[
\tilde{z}_{n}^{P}\right]  =\left[  \tilde{z}_{1}^{P}\right]  =\tilde{z}%
_{10}^{P}=1-\left\vert -\frac{k_{1}^{P}}{k_{3}^{P}}\right\vert ,\nonumber\\
\tilde{Z}_{n}^{L}  &  =\left[  \tilde{z}_{1}^{L}\right]  ,\cdots,\left[
\tilde{z}_{n}^{L}\right]  =\left[  \tilde{z}_{1}^{L}\right]  =\tilde{z}%
_{10}^{L}=1-\left\vert -\frac{k_{1}^{L}}{k_{3}^{L}}\right\vert
\end{align}
and the order $K$ in Equation (\ref{kk}) is
\begin{align}
\text{ }K  &  =\underset{\{\text{for all }r_{j}^{T}\neq0\}}{\sum
j}+\underset{\{\text{for all }r_{j}^{P}\neq0\}}{\sum j}+\underset{\{\text{for
all }r_{j}^{L}\neq0\}}{\sum j}\nonumber\\
&  =1+1+1=3.
\end{align}

\subsubsection{Example Two}

We take the tensor state to be%
\begin{equation}
\left\vert \text{state}\right\rangle =\left(  \alpha_{-1}^{T}\right)
^{r_{1}^{T}}\left(  \alpha_{-2}^{T}\right)  ^{r_{2}^{T}}\left(  \alpha
_{-5}^{T}\right)  ^{r_{5}^{T}}\left(  \alpha_{-6}^{T}\right)  ^{r_{6}^{T}%
}|0,k\rangle.
\end{equation}

The LSSA in Equation (\ref{st1}) can be calculated to be
\begin{align}
A_{st}^{(r_{1}^{T},r_{1}^{P},r_{l}^{L})}  &  =\left(  -k_{3}^{T}\right)
^{r_{1}^{T}}\left(  -k_{3}^{T}\right)  ^{r_{2}^{T}}\left(  -4!k_{3}%
^{T}\right)  ^{r_{5}^{T}}\left(  -5!k_{3}^{T}\right)  ^{r_{6}^{T}}B\left(
-\frac{t}{2}-1,-\frac{s}{2}-1\right) \nonumber\\
&  \cdot F_{D}^{(14)}\left(
\begin{array}
[c]{c}%
-\frac{t}{2}-1;-r_{1}^{T},\underset{2}{\underbrace{-r_{2}^{T},-r_{2}^{T}}%
},\underset{5}{\underbrace{-r_{5}^{T},-r_{5}^{T},-r_{5}^{T},-r_{5}^{T}%
,-r_{5}^{T}}},\underset{6}{\underbrace{-r_{6}^{T},-r_{6}^{T},-r_{6}^{T}%
,-r_{6}^{T},-r_{6}^{T},-r_{6}^{T}}};\\
\frac{u}{2}+2-N;\tilde{z}_{10}^{T},\underset{2}{\underbrace{\tilde{z}_{20}%
^{T},\tilde{z}_{21}^{T}}},\underset{5}{\underbrace{\tilde{z}_{50}^{T}%
,\tilde{z}_{51}^{T},\tilde{z}_{52}^{T},\tilde{z}_{53}^{T},\tilde{z}_{54}^{T}}%
},\underset{6}{\underbrace{\tilde{z}_{60}^{T},\tilde{z}_{61}^{T},\tilde
{z}_{62}^{T},\tilde{z}_{63}^{T},\tilde{z}_{64}^{T},\tilde{z}_{65}^{T}}}%
\end{array}
\right)
\end{align}
where the arguments in $F_{D}^{(14)}$ are calculated to be

\begin{align}
R_{n}^{T}  &  =\left\{  -r_{1}^{T}\right\}  ^{1},\cdots,\left\{  -r_{n}%
^{T}\right\}  ^{k}=\left\{  -r_{1}^{T}\right\}  ^{1},\left\{  -r_{2}%
^{T}\right\}  ^{2},\left\{  -r_{5}^{T}\right\}  ^{5},\left\{  -r_{6}%
^{T}\right\}  ^{6}\nonumber\\
&  =-r_{1}^{T},\underset{2}{\underbrace{-r_{2}^{T},-r_{2}^{T}}}%
,\underset{5}{\underbrace{-r_{5}^{T},-r_{5}^{T},-r_{5}^{T},-r_{5}^{T}%
,-r_{5}^{T}}},\underset{6}{\underbrace{-r_{6}^{T},-r_{6}^{T},-r_{6}^{T}%
,-r_{6}^{T},-r_{6}^{T},-r_{6}^{T}}}\\
\tilde{Z}_{n}^{T}  &  =\left[  \tilde{z}_{1}^{T}\right]  ,\cdots,\left[
\tilde{z}_{n}^{T}\right]  =\left[  \tilde{z}_{1}^{T}\right]  ,\left[
\tilde{z}_{2}^{T}\right]  ,\left[  \tilde{z}_{5}^{T}\right]  ,\left[
\tilde{z}_{6}^{T}\right] \nonumber\\
&  =\tilde{z}_{10}^{T},\underset{2}{\underbrace{\tilde{z}_{20}^{T},\tilde
{z}_{21}^{T}}},\underset{5}{\underbrace{\tilde{z}_{50}^{T},\tilde{z}_{51}%
^{T},\tilde{z}_{52}^{T},\tilde{z}_{53}^{T},\tilde{z}_{54}^{T}}}%
,\underset{6}{\underbrace{\tilde{z}_{60}^{T},\tilde{z}_{61}^{T},\tilde{z}%
_{62}^{T},\tilde{z}_{63}^{T},\tilde{z}_{64}^{T},\tilde{z}_{65}^{T}}}%
\end{align}
and%
\begin{align}
\text{ }K  &  =\underset{\{\text{for all }r_{j}^{T}\neq0\}}{\sum
j}+\underset{\{\text{for all }r_{j}^{P}\neq0\}}{\sum j}+\underset{\{\text{for
all }r_{j}^{L}\neq0\}}{\sum j}\nonumber\\
&  =\left(  1+2+5+6\right)  +0+0=14.
\end{align}

In the following subsections, we discuss the exact $SL(K+3,C)$ symmetry of the
LSSA. For simplicity, we begin with the simple $SL(4,C)$ symmetry with
$K=1.$

\subsection{The $SL(4,C)$ Symmetry}

In this section, for illustration, we first consider the simplest $K=1$ case
with $SL(4,C)$ symmetry. For a given $K$, there can be LSSAs with different
mass levels $N$. As an example, for the case of $K=1$, there are three types of
LSSA:
\begin{align}
(\alpha_{-1}^{T})^{p_{1}}\text{ , }F_{D}^{(1)}\left(  -\frac{t}{2}%
-1,-p_{1},,\frac{u}{2}+2-p_{1},1\right)  \text{ , }N  &  =p_{1},\nonumber\\
(\alpha_{-1}^{P})^{q_{1}}\text{ , }F_{D}^{(1)}\left(  -\frac{t}{2}%
-1,-q_{1},\frac{u}{2}+2-q_{1},\left[  \tilde{z}_{1}^{P}\right]  \right)
\text{ , }N  &  =q_{1},\nonumber\\
(\alpha_{-1}^{L})^{r_{1}}\text{ , }F_{D}^{(1)}\left(  -\frac{t}{2}%
-1,-r_{1},\frac{u}{2}+2-r_{1},\left[  \tilde{z}_{1}^{L}\right]  \right)
\text{ , }N  &  =r_{1}. \label{rel}%
\end{align}
To calculate the group representation of the LSSA for $K=1$, we define
\cite{slkc}%
\begin{equation}
f_{ac}^{b}\left(  \alpha;\beta;\gamma;x\right)  =B\left(  \gamma-\alpha
,\alpha\right)  F_{D}^{\left(  1\right)  }\left(  \alpha;\beta;\gamma
;x\right)  a^{\alpha}b^{\beta}c^{\gamma}. \label{id}%
\end{equation}

We see that the LSSA in Equation (\ref{st1}) for the case of $K=1$ corresponds to
the case $a=1=c$, and can be written as%
\begin{equation}
A_{st}^{R^{X}}=f_{11}^{-k_{3}^{X}}\left(  -\frac{t}{2}-1;R^{X};\frac{u}%
{2}+2-N;\tilde{Z}^{X}\right)  .
\end{equation}

We  can now introduce the $(K+3)^{2}-1$ $=(1+3)^{2}-1=15$ generators of
$SL(4,C)$ \mbox{group \cite{sl4c,slkc}}%
\begin{align}
E_{\alpha}  &  =a\left(  x\partial_{x}+a\partial_{a}\right)  ,\nonumber\\
E_{-\alpha}  &  =\frac{1}{a}\left[  x\left(  1-x\right)  \partial
_{x}+c\partial_{c}-a\partial_{a}-xb\partial_{b}\right]  ,\nonumber\\
E_{\beta}  &  =b\left(  x\partial_{x}+b\partial_{b}\right)  ,\nonumber\\
E_{-\beta}  &  =\frac{1}{b}\left[  x\left(  1-x\right)  \partial_{x}%
+c\partial_{c}-b\partial_{b}-xa\partial_{a}\right]  ,\nonumber\\
E_{\gamma}  &  =c\left[  \left(  1-x\right)  \partial_{x}+c\partial
_{c}-a\partial_{a}-b\partial_{b}\right]  ,\nonumber\\
E_{-\gamma}  &  =-\frac{1}{c}\left(  x\partial_{x}+c\partial_{c}-1\right)
,\nonumber\\
E_{\beta\gamma}  &  =bc\left[  \left(  x-1\right)  \partial_{x}+b\partial
_{b}\right]  ,\nonumber\\
E_{-\beta,-\gamma}  &  =\frac{1}{bc}\left[  x\left(  x-1\right)  \partial
_{x}+xa\partial_{a}-c\partial_{c}+1\right]  ,\nonumber\\
E_{\alpha\gamma}  &  =ac\left[  \left(  1-x\right)  \partial_{x}-a\partial
_{a}\right]  ,\nonumber\\
E_{-\alpha,-\gamma}  &  =\frac{1}{ac}\left[  x\left(  1-x\right)  \partial
_{x}-xb\partial_{b}+c\partial_{c}-1\right]  ,\nonumber\\
E_{\alpha\beta\gamma}  &  =abc\partial_{x},\nonumber\\
E_{-\alpha,-\beta,-\gamma}  &  =\frac{1}{abc}\left[  x\left(  x-1\right)
\partial_{x}-c\partial_{c}+xb\partial_{b}+xa\partial_{a}-x+1\right]
,\nonumber\\
J_{\alpha}  &  =a\partial_{a},\nonumber\\
J_{\beta}  &  =b\partial_{b},\nonumber\\
J_{\gamma}  &  =c\partial_{c}, \label{def}%
\end{align}
and calculate their operations on the basis of functions \cite{sl4c,slkc}%
\begin{align}
E_{\alpha}f_{ac}^{b}\left(  \alpha;\beta;\gamma;x\right)   &  =\left(
\gamma-\alpha-1\right)  f_{ac}^{b}\left(  \alpha+1;\beta;\gamma;x\right)
,\nonumber\\
E_{\beta}f_{ac}^{b}\left(  \alpha;\beta;\gamma;x\right)   &  =\beta f_{ac}%
^{b}\left(  \alpha;\beta+1;\gamma;x\right)  ,\nonumber\\
E_{\gamma}f_{ac}^{b}\left(  \alpha;\beta;\gamma;x\right)   &  =\left(
\gamma-\beta\right)  f_{ac}^{b}\left(  \alpha;\beta;\gamma+1;x\right)
,\nonumber\\
E_{\beta\gamma}f_{ac}^{b}\left(  \alpha;\beta;\gamma;x\right)   &  =\beta
f_{ac}^{b}\left(  \alpha;\beta+1;\gamma+1;x\right)  ,\nonumber\\
E_{\alpha\gamma}f_{ac}^{b}\left(  \alpha;\beta;\gamma;x\right)   &  =\left(
\beta-\gamma\right)  f_{ac}^{b}\left(  \alpha+1;\beta;\gamma+1;x\right)
,\nonumber\\
E_{\alpha\beta\gamma}f_{ac}^{b}\left(  \alpha;\beta;\gamma;x\right)   &
=\beta f_{ac}^{b}\left(  \alpha+1;\beta+1;\gamma+1;x\right)  ,\nonumber\\
E_{-\alpha}f_{ac}^{b}\left(  \alpha;\beta;\gamma;x\right)   &  =\left(
\alpha-1\right)  f_{ac}^{b}\left(  \alpha-1;\beta;\gamma;x\right)
,\nonumber\\
E_{-\beta}f_{ac}^{b}\left(  \alpha;\beta;\gamma;x\right)   &  =\left(
\gamma-\beta\right)  f_{ac}^{b}\left(  \alpha;\beta-1;\gamma;x\right)
,\nonumber\\
E_{-\gamma}f_{ac}^{b}\left(  \alpha;\beta;\gamma;x\right)   &  =\left(
\alpha+1-\gamma\right)  f_{ac}^{b}\left(  \alpha;\beta;\gamma-1;x\right)
,\nonumber\\
E_{-\beta,-\gamma}f_{ac}^{b}\left(  \alpha;\beta;\gamma;x\right)   &  =\left(
\alpha-\gamma+1\right)  f_{ac}^{b}\left(  \alpha;\beta-1;\gamma-1;x\right)
,\nonumber\\
E_{-\alpha,-\gamma}f_{ac}^{b}\left(  \alpha;\beta;\gamma;x\right)   &
=\left(  \alpha-1\right)  f_{ac}^{b}\left(  \alpha-1;\beta;\gamma-1;x\right)
,\nonumber\\
E_{-\alpha,-\beta,-\gamma}f_{ac}^{b}\left(  \alpha;\beta;\gamma;x\right)   &
=\left(  -\alpha+1\right)  f_{ac}^{b}\left(  \alpha-1;\beta-1;\gamma
-1;x\right)  ,\nonumber\\
J_{\alpha}f_{ac}^{b}\left(  \alpha;\beta;\gamma;x\right)   &  =\alpha
f_{ac}^{b}\left(  \alpha;\beta;\gamma;x\right)  ,\nonumber\\
J_{\beta}f_{ac}^{b}\left(  \alpha;\beta;\gamma;x\right)   &  =\beta f_{ac}%
^{b}\left(  \alpha;\beta;\gamma;x\right)  ,\nonumber\\
J_{\gamma}f_{ac}^{b}\left(  \alpha;\beta;\gamma;x\right)   &  =\gamma
f_{ac}^{b}\left(  \alpha;\beta;\gamma;x\right)  . \label{op}%
\end{align}
It is important to note, for example, that since $\beta$ is a nonpositive
integer, the operation by $E_{-\beta}$ will not be terminated as in the case
of the finite dimensional representation of a compact Lie group. Here the
representation is infinite-dimensional. On the other hand, a simple
calculation gives%
\vspace{6pt}
\begin{align*}
\left[  E_{\alpha},E_{-\alpha}\right]   &  =2J_{\alpha}-J_{\gamma},\\
\left[  E_{\beta},E_{-\beta}\right]   &  =2J_{\beta}-J_{\gamma},\\
\left[  E_{\gamma},E_{-\gamma}\right]   &  =2J_{\gamma}-\left(  J_{\alpha
}+J_{\beta}+1\right)  ,
\end{align*}
which suggests the Cartan subalgebra%
\begin{equation}
\left[  J_{\alpha},J_{\beta}\right]  =0,\left[  J_{\beta},J_{\gamma}\right]
=0,\left[  J_{\alpha},J_{\gamma}\right]  =0. \label{cartan}%
\end{equation}

Indeed, if we redefine%
\begin{align*}
J_{\alpha}^{\prime}  &  =J_{\alpha}-\frac{1}{2}J_{\gamma},\\
J_{\beta}^{\prime}  &  =J_{\beta}-\frac{1}{2}J_{\gamma},\\
J_{\gamma}^{\prime}  &  =J_{\gamma}-\frac{1}{2}\left(  J_{\alpha}+J_{\beta
}+1\right)  ,
\end{align*}
we discover that each of the following six triplets \cite{sl4c,slkc}%
\begin{align*}
&  \left\{  J^{+},J^{-},J^{0}\right\}  \equiv\left\{  E_{\alpha},E_{-\alpha
},J_{\alpha}^{\prime}\right\}  ,\left\{  E_{\beta},E_{-\beta},J_{\beta
}^{\prime}\right\}  ,\\
&  \left\{  E_{\gamma},E_{-\gamma},J_{\gamma}^{\prime}\right\}  ,\left\{
E_{\alpha,\beta,\gamma},E_{-\alpha,-\beta,-\gamma},J_{\alpha}^{\prime
}+J_{\beta}^{\prime}+J_{\gamma}^{\prime}\right\}  ,\\
&  \left\{  E_{\alpha\gamma},E_{-\alpha,-\gamma},J_{\alpha}^{\prime}%
+J_{\gamma}^{\prime}\right\}  ,\left\{  E_{\alpha\beta},E_{-\alpha,-\beta
},J_{\alpha}^{\prime}+J_{\beta}^{\prime}\right\}
\end{align*}
constitutes the well-known commutation relations%
\begin{equation}
\left[  J^{0},J^{\pm}\right]  =\pm J^{\pm},\left[  J^{+},J^{-}\right]
=2J^{0}. \label{com}%
\end{equation}

\subsection{The General SL($K+3$,C) Symmetry}

We are now ready to generalize the calculation of the previous section and
calculate the group representation of the LSSA for general $K$. We first
define \cite{slkc}
\begin{align}
&  f_{ac}^{b_{1}\cdots b_{K}}\left(  \alpha;\beta_{1},\cdots,\beta_{K}%
;\gamma;x_{1},\cdots,x_{K}\right) \nonumber\\
&  =B\left(  \gamma-\alpha,\alpha\right)  F_{D}^{\left(  K\right)  }\left(
\alpha;\beta_{1},\cdots,\beta_{K};\gamma;x_{1},\cdots,x_{K}\right)  a^{\alpha
}b_{1}^{\beta_{1}}\cdots b_{K}^{\beta_{K}}c^{\gamma}. \label{id2}%
\end{align}

Note that the LSSA in Equation (\ref{st1}) corresponds to the case $a=1=c$ and can
be written~as
\begin{equation}
A_{st}^{(r_{n}^{T},r_{m}^{P},r_{l}^{L})}=f_{11}^{-(n-1)!k_{3}^{T}%
,-(m-1)!k_{3}^{P},-(l-1)!k_{3}^{L}}\left(  -\frac{t}{2}-1;R_{n}^{T},R_{m}%
^{P},R_{l}^{L};\frac{u}{2}+2-N;\tilde{Z}_{n}^{T},\tilde{Z}_{m}^{P},\tilde
{Z}_{l}^{L}\right).
\end{equation}
It is possible to extend the calculation of the $SL(4,%
\mathbb{C}
)$ symmetry group for the $K=1$ case discussed in the previous section to the
general $SL(K+3,%
\mathbb{C}
)$ group. We first introduce the $(K+3)^{2}-1$ generators of the $SL(K+3,C)$ group
($k=1,2,...K$) \cite{sl4c,slkc}%
\vspace{6pt}
\begin{align}
E^{\alpha}  &  =a\left(  \underset{j}{%
{\displaystyle\sum}
}x_{j}\partial_{j}+a\partial_{a}\right)  ,\nonumber\\
E^{\beta_{k}}  &  =b_{k}\left(  x_{k}\partial_{k}+b_{k}\partial_{b_{k}%
}\right)  ,\nonumber\\
E^{\gamma}  &  =c\left(  \underset{j}{%
{\displaystyle\sum}
}\left(  1-x_{j}\right)  \partial_{x_{j}}+c\partial_{c}-a\partial
_{a}-\underset{j}{%
{\displaystyle\sum}
}b_{j}\partial_{b_{j}}\right)  ,\nonumber\\
E^{\alpha\gamma}  &  =ac\left(  \underset{j}{%
{\displaystyle\sum}
}\left(  1-x_{j}\right)  \partial_{x_{j}}-a\partial_{a}\right)  ,\nonumber\\
E^{\beta_{k}\gamma}  &  =b_{k}c\left[  \left(  x_{k}-1\right)  \partial
_{x_{k}}+b_{k}\partial_{b_{k}}\right]  ,\nonumber\\
E^{\alpha\beta_{k}\gamma}  &  =ab_{k}c\partial_{x_{k}},\nonumber\\
E_{\alpha}  &  =\frac{1}{a}\left[  \underset{j}{%
{\displaystyle\sum}
}x_{j}\left(  1-x_{j}\right)  \partial_{x_{j}}+c\partial_{c}-a\partial
_{a}-\underset{j}{%
{\displaystyle\sum}
}x_{j}b_{j}\partial_{b_{j}}\right]  ,\nonumber\\
E_{\beta_{k}}  &  =\frac{1}{b_{k}}\left[  x_{k}\left(  1-x_{k}\right)
\partial_{x_{k}}+x_{k}\underset{j\neq k}{%
{\displaystyle\sum}
}\left(  1-x_{j}\right)  x_{j}\partial_{x_{j}}+c\partial_{c}-x_{k}%
a\partial_{a}-\underset{j}{%
{\displaystyle\sum}
}b_{j}\partial_{u_{j}}\right]  ,\nonumber\\
E_{\gamma}  &  =-\frac{1}{c}\left(  \underset{j}{%
{\displaystyle\sum}
}x_{j}\partial_{x_{j}}+c\partial_{c}-1\right)  ,\nonumber\\
E_{\alpha\gamma}  &  =\frac{1}{ac}\left[  \underset{j}{%
{\displaystyle\sum}
}x_{j}\left(  1-x_{j}\right)  \partial_{x_{j}}-\underset{j}{%
{\displaystyle\sum}
}x_{j}b_{j}\partial_{b_{j}}+c\partial_{c}-1\right]  ,\nonumber\\
E_{\beta_{k}\gamma}  &  =\frac{1}{b_{k}c}\left[  x_{k}\left(  x_{k}-1\right)
\partial_{x_{k}}+\underset{j\neq k}{%
{\displaystyle\sum}
}\left(  x_{j}-1\right)  x_{j}\partial_{x_{j}}+x_{k}a\partial_{a}%
-c\partial_{c}+1\right]  ,\nonumber\\
E_{\alpha\beta_{k}\gamma}  &  =\frac{1}{ab_{k}c}\left[  \underset{j}{%
{\displaystyle\sum}
}x_{j}\left(  x_{j}-1\right)  \partial_{x_{j}}-c\partial_{c}+x_{k}%
a\partial_{a}+\underset{j}{%
{\displaystyle\sum}
}x_{j}b_{j}\partial_{b_{j}}-x_{k}+1\right]  ,\nonumber\\
E_{\beta_{p}}^{\beta_{k}}  &  =\frac{b_{k}}{b_{p}}\left[  \left(  x_{k}%
-x_{p}\right)  \partial_{z_{k}}+b_{k}\partial_{b_{k}}\right]  ,(k\neq
p),\nonumber\\
J_{\alpha}  &  =a\partial_{a},\nonumber\\
J_{\beta_{k}}  &  =b_{k}\partial_{b_{k}},\nonumber\\
J_{\gamma}  &  =c\partial_{c}. \label{def2}%
\end{align}

Note that we have used the upper indices to denote the ``raising operators'' and
the lower indices to denote the ``lowering operators''. The number of generators
can be counted in the following way. There are $1$ $E^{\alpha}$, $K$
$E^{\beta_{k}}$, $1$ $E^{\gamma}$,$1$ $E^{\alpha\gamma}$,$K$ $E^{\beta
_{k}\gamma}$ and $K$ $E^{\alpha\beta_{k}\gamma}$ which sum up to $3K+3$
raising generators. There are also $3K+3$ lowering operators. In addition,
there are $K\left(  K-1\right)  $ $E_{\beta_{p}}^{\beta_{k}}$ and $K+2$
$\ $\ $\ J$, corresponding to the Cartan subalgebra. In summary, the total number of generators
is $2(3K+3)+K(K-1)+K+2=(K+3)^{2}-1$. It is straightforward to calculate the
operation of these generators on the basis of functions ($k=1,2,\ldots, K$)
\cite{slkc}%

\begin{align}
E^{\alpha}f_{ac}^{b_{1}\cdots b_{K}}\left(  \alpha\right)   &  =\left(
\gamma-\alpha-1\right)  f_{ac}^{b_{1}\cdots b_{K}}\left(  \alpha+1\right)
,\nonumber\\
E^{\beta_{k}}f_{ac}^{b_{1}\cdots b_{K}}\left(  \beta_{k}\right)   &
=\beta_{k}f_{ac}^{b_{1}\cdots b_{K}}\left(  \beta_{k}+1\right)  ,\nonumber\\
E^{\gamma}f_{ac}^{b_{1}\cdots b_{K}}\left(  \gamma\right)   &  =\left(
\gamma-\underset{j}{%
{\displaystyle\sum}
}\beta_{j}\right)  f_{ac}^{b_{1}\cdots b_{K}}\left(  \gamma+1\right)
,\nonumber\\
E^{\alpha\gamma}f_{ac}^{b_{1}\cdots b_{K}}\left(  \alpha;\gamma\right)   &
=\left(  \underset{j}{%
{\displaystyle\sum}
}\beta_{j}-\gamma\right)  f_{ac}^{b_{1}\cdots b_{K}}\left(  \alpha
+1;\gamma+1\right)  ,\nonumber\\
E^{\beta_{k}\gamma}f_{ac}^{b_{1}\cdots b_{K}}\left(  \beta_{k};\gamma\right)
&  =\beta_{k}f_{ac}^{b_{1}\cdots b_{K}}\left(  \beta_{k}+1;\gamma+1\right)
,\nonumber\\
E^{\alpha\beta_{k}\gamma}f_{ac}^{b_{1}\cdots b_{K}}\left(  \alpha;\beta
_{k};\gamma\right)   &  =\beta_{k}f_{ac}^{b_{1}\cdots b_{K}}\left(
\alpha+1;\beta_{k}+1;\gamma+1\right)  ,\nonumber\\
E_{\alpha}f_{ac}^{b_{1}\cdots b_{K}}\left(  \alpha\right)   &  =\left(
\alpha-1\right)  f_{ac}^{b_{1}\cdots b_{K}}\left(  \alpha-1\right)
,\nonumber\\
E_{\beta_{k}}f_{ac}^{b_{1}\cdots b_{K}}\left(  \beta_{k}\right)   &  =\left(
\gamma-\underset{j}{%
{\displaystyle\sum}
}\beta_{j}\right)  f_{ac}^{b_{1}\cdots b_{K}}\left(  \beta_{k}-1\right)
,\nonumber\\
E_{\gamma}f_{ac}^{b_{1}\cdots b_{K}}\left(  \gamma\right)   &  =\left(
\alpha-\gamma+1\right)  f_{ac}^{b_{1}\cdots b_{K}}\left(  \gamma-1\right)
,\nonumber\\
E_{\alpha\gamma}f_{ac}^{b_{1}\cdots b_{K}}\left(  \alpha;\gamma\right)   &
=\left(  \alpha-1\right)  f_{ac}^{b_{1}\cdots b_{K}}\left(  \alpha
-1;\gamma-1\right)  ,\nonumber\\
E_{\beta_{k}\gamma}f_{ac}^{b_{1}\cdots b_{K}}\left(  \beta_{k};\gamma\right)
&  =\left(  \alpha-\gamma+1\right)  f_{ac}^{b_{1}\cdots b_{K}}\left(
\beta_{k}-1;\gamma-1\right)  ,\nonumber\\
E_{\alpha\beta_{k}\gamma}f_{ac}^{b_{1}\cdots b_{K}}\left(  \alpha;\beta
_{k};\gamma\right)   &  =\left(  1-\alpha\right)  f_{ac}^{b_{1}\cdots b_{K}%
}\left(  \alpha-1;\beta_{k}-1;\gamma-1\right)  ,\nonumber\\
E_{\beta_{p}}^{\beta_{k}}f_{ac}^{b_{1}\cdots b_{K}}\left(  \beta_{k};\beta
_{p}\right)   &  =\beta_{k}f_{ac}^{b_{1}\cdots b_{K}}\left(  \beta_{k}%
+1;\beta_{p}-1\right)  ,\nonumber\\
J_{\alpha}f_{ac}^{b_{1}\cdots b_{K}}\left(  \alpha;\beta_{k};\gamma\right)
&  =\alpha f_{ac}^{b_{1}\cdots b_{K}}\left(  \alpha;\beta_{k};\gamma\right)
,\nonumber\\
J_{\beta_{k}}f_{ac}^{b_{1}\cdots b_{K}}\left(  \alpha;\beta_{k};\gamma\right)
&  =\beta_{k}f_{ac}^{b_{1}\cdots b_{K}}\left(  \alpha;\beta_{k};\gamma\right)
,\nonumber\\
J_{\gamma}f_{ac}^{b_{1}\cdots b_{K}}\left(  \alpha;\beta_{k};\gamma\right)
&  =\gamma f_{ac}^{b_{1}\cdots b_{K}}\left(  \alpha;\beta_{k};\gamma\right)
\label{op2}%
\end{align}
where, for simplicity, we have omitted those arguments in $f_{ac}^{b_{1}\cdots
b_{K}}$ that remain the same after the operation. The commutation relations
of the $SL(K+3)$ Lie algebra can be calculated in the following way. In
addition to the Cartan subalgebra for the $K+2$ generators $\left\{
J_{\alpha},J_{\beta_{k}},J_{\gamma}\right\}  $, we redefine%
\begin{align}
J_{\alpha}^{\prime}  &  =J_{\alpha}-\frac{1}{2}J_{\gamma},\nonumber\\
J_{\beta_{k}}^{\prime}  &  =J_{\beta_{k}}-\frac{1}{2}J_{\gamma}%
+\underset{j\neq k}{\sum}J_{\beta_{j}},\nonumber\\
J_{\gamma}^{\prime}  &  =J_{\gamma}-\frac{1}{2}\left(  J_{\alpha
}+\underset{j}{\sum}J_{\beta_{j}}+1\right)  .
\end{align}

We discover that each of the following seven triplets \cite{slkc}%
\begin{align}
&  \left\{  J^{+},J^{-},J^{0}\right\}  \equiv\left\{  E^{\alpha},E_{\alpha
},J_{\alpha}^{\prime}\right\}  ,\left\{  E^{\beta_{k}},E_{\beta_{k}}%
,J_{\beta_{k}}^{\prime}\right\}  ,\nonumber\\
&  \left\{  E^{\gamma},E_{\gamma},J_{\gamma}^{\prime}\right\}  ,\left\{
E^{\alpha\beta_{k}\gamma},E_{\alpha\beta_{k}\gamma},J_{\alpha}^{\prime
}+J_{\beta_{k}}^{\prime}+J_{\gamma}^{\prime}\right\}  ,\nonumber\\
&  \left\{  E^{\alpha\gamma},E_{\alpha\gamma},J_{\alpha}^{\prime}+J_{\gamma
}^{\prime}\right\}  ,\left\{  E^{\alpha\beta_{k}},E_{\alpha\beta_{k}%
},J_{\alpha}^{\prime}+J_{\beta_{k}}^{\prime}\right\}  ,\nonumber\\
&  \left\{  E_{\beta_{p}}^{\beta_{l}},E_{\beta_{l}}^{\beta_{p}},J_{\beta_{l}%
}^{\prime}-J_{\beta_{p}}^{\prime}\right\}  \label{33}%
\end{align}
satisfies the commutation relations in Equation (\ref{com}).

Finally, in addition to Equation (\ref{33}), there is another compact way to write the Lie algebra commutation relations of $SL(K+3,C).$ Indeed, one can
check that the Lie algebra commutation relations of $SL(K+3,C)$ can be written
as \cite{slkc}
\begin{equation}
\left[  \mathcal{E}_{ij},\mathcal{E}_{kl}\right]  =\delta_{jk}\mathcal{E}%
_{il}-\delta_{li}\mathcal{E}_{kj}%
\end{equation}
with the following identifications:
\begin{align}
E^{\alpha}  &  =\mathcal{E}_{12},E_{\alpha}=\mathcal{E}_{21},E^{\beta_{k}%
}=\mathcal{E}_{k+3,3},E_{\beta}=\mathcal{E}_{3,k+3},\nonumber\\
E^{\gamma}  &  =\mathcal{E}_{31},E_{\gamma}=\mathcal{E}_{13},E^{\alpha\gamma
}=\mathcal{E}_{32},E_{\alpha\gamma}=\mathcal{E}_{23},\nonumber\\
E^{\beta_{k}\gamma}  &  =-\mathcal{E}_{k+3,1},E_{\beta_{k}\gamma}%
=-\mathcal{E}_{1,k+3},E_{\alpha\beta_{k}\gamma}=-\mathcal{E}_{k+3,2}%
,\nonumber\\
E_{\alpha\beta_{k}\gamma}  &  =-\mathcal{E}_{2,k+3},J_{\alpha}^{\prime}%
=\frac{1}{2}\left(  \mathcal{E}_{11}-\mathcal{E}_{22}\right)  ,J_{\beta_{k}%
}^{\prime}=\frac{1}{2}\left(  \mathcal{E}_{k+3,k+3}-\mathcal{E}_{33}\right)
,J_{\gamma}^{\prime}=\frac{1}{2}\left(  \mathcal{E}_{33}-\mathcal{E}%
_{11}\right)  .
\end{align}

\subsection{Discussion}

There are some special properties in the $SL(K+3,%
\mathbb{C}
)$ group representation of the LSSA that make it different from the usual
symmetry group representation of a physical system. First, the set of LSSA
does not fill up the whole representation space $V$. For example, states
$f_{ac}^{b_{1}\cdots b_{K}}\left(  \alpha;\beta_{1},\cdots,\beta_{K}%
;\gamma;x_{1},\cdots,x_{K}\right)  $ in $V$ with $a\neq1$ or $c\neq1$ are not LSSAs.

Indeed, there are more states in $V$ with $K\geq2$ that are not LSSAs either.
We give one example in the following. For $K=2$, there are six types of LSSAs:
$(\omega=-1)$%
\begin{align}
(\alpha_{-1}^{T})^{p_{1}}(\alpha_{-1}^{P})^{q_{1}}\text{,}F_{D}^{(2)}%
(a,-p_{1},-q_{1},c-p_{1}-q_{1},1,\left[  \tilde{z}_{1}^{P}\right]  )\text{,}N
&  =p_{1}+q_{1},\\
(\alpha_{-1}^{T})^{p_{1}}(\alpha_{-1}^{L})^{r_{1}}\text{,}F_{D}^{(2)}%
(a,-p_{1},-r_{1},c-p_{1}-r_{1},1,\left[  \tilde{z}_{1}^{L}\right]  )\text{,}N
&  =p_{1}+r_{1},\\
(\alpha_{-1}^{P})^{q_{1}}(\alpha_{-1}^{L})^{r_{1}}\text{,}F_{D}^{(2)}%
(a,-q_{1},-r_{1},c-q_{1}-r_{1},\left[  \tilde{z}_{1}^{P}\right]  ,\left[
\tilde{z}_{1}^{L}\right]  )\text{,}N  &  =q_{1}+r_{1},\\
(\alpha_{-2}^{T})^{p_{2}}\text{ , }F_{D}^{(2)}(a,-p_{2},-p_{2},c-2p_{2}%
,1,1)\text{ , }N  &  =2p_{2},\label{2a}\\
(\alpha_{-2}^{P})^{q_{2}}\text{ , }F_{D}^{(2)}(a,-q_{2},-q_{2},c-2q_{2}%
,1-z_{2}^{P},1-\omega z_{2}^{P}),N  &  =2q_{2},\label{2b}\\
(\alpha_{-2}^{L})^{r_{2}}\text{ , }F_{D}^{(2)}(a,-r_{2},-r_{2},c-2r_{2}%
,1-z_{2}^{L},1-\omega z_{2}^{L}),N  &  =2r_{2}. \label{2c}%
\end{align}
One can show that those states obtained from the operation by $E_{\beta}$ in
either states in \mbox{Equations (\ref{2a})--(\ref{2c})} are not LSSAs. However, it is shown in Section 3 that all states in $V$, including those ``auxiliary states''
which are not LSSAs as stated above, can be exactly solved by recurrence relations
or the $SL(K+3,%
\mathbb{C}
)$ group and expressed in terms of one amplitude. These ``auxiliary states''
and states with $a\neq1$ or $c\neq1$ in $V$ may represent other SSAs---e.g., SSAs
of two tachyons and two arbitrary string states, etc.---which will be considered
in the near future.%


\section{Solving LSSA through Recurrence Relations}

In the previous section, the string scattering amplitudes of three tachyons
and one arbitrary string states in the 26D open bosonic string theory  were
obtained in terms of the $D$-type Lauricella functions; i.e., the LSSA in
Equation (\ref{st1}). The symmetry of the LSSA was also discussed by constructing
the $SL(K+3,%
\mathbb{C}
)$ group for the $D$-type Lauricella functions $F_{D}^{(K)}\left(
\alpha;\beta_{1},...,\beta_{K};\gamma;x_{1},...,x_{K}\right)  $. It is natural
to suspect that the LSSAs are dependent on each other due to the symmetry between
them. In fact, we are able to show that all the LSSAs are related to a single
LSSA by the recurrence relations of the $D$-type Lauricella functions.

To solve all the LSSAs, a key observation is that all arguments $\beta_{m}$ in
the Lauricella functions $F_{D}^{(K)}\left(  \alpha;\beta_{1},...,\beta
_{K};\gamma;x_{1},...,x_{K}\right)  $ in the LSSA (\ref{st1}) are nonpositive
integers. We show that this plays a key role in proving the solvability of
all the LSSAs below.

The generalization of the $2+2$ recurrence relations of the Appell functions
to the $K+2$ recurrence relations of the Lauricella functions was given in
\cite{1707.01281}. One can use these $K+2$ recurrence relations to reduce all
the Lauricella functions $F_{D}^{(K)}$ in the LSSA (\ref{st1}) to the Gauss
hypergeometry functions $_{2}F_{1}(\alpha,\beta,\gamma)$. Then, all the LSSAs
can be solved by deriving a multiplication theorem for the Gauss hypergeometry functions.

In this section, we will review the steps presented in \cite{1707.01281}.

\subsection{Recurrence Relations of the LSSA}

For $K=2$, the Lauricella functions $D$-type $F_{D}^{(K)}\left(  \alpha
;\beta_{1},...,\beta_{K};\gamma;x_{1},...,x_{K}\right)  $ reduce to the
type-$1$Appell functions $F_{1}\left(  \alpha;b_{1},\beta_{2};\gamma
,x,y\right)  $. The four fundamental recurrence relations which link the
contiguous functions are%
\begin{align}
\left(  \alpha-\beta_{1}-\beta_{2}\right)  F_{1}\left(  \alpha;\beta_{1}%
,\beta_{2};\gamma,x,y\right)  -\alpha F_{1}\left(  \alpha+1;\beta_{1}%
,\beta_{2};\gamma,x,y\right)   & \nonumber\\
+\beta_{1}F_{1}\left(  \alpha;\beta_{1}+1,\beta_{2};\gamma,x,y\right)
+\beta_{2}F_{1}\left(  \alpha;\beta_{1},\beta_{2}+1;\gamma,x,y\right)   &
=0,\\
\gamma F_{1}\left(  \alpha;\beta_{1},\beta_{2};\gamma,x,y\right)  -\left(
\gamma-\alpha\right)  F_{1}\left(  \alpha;\beta_{1},\beta_{2};\gamma
+1,x,y\right)   & \nonumber\\
-\alpha F_{1}\left(  \alpha+1;\beta_{1},\beta_{2};\gamma+1,x,y\right)   &
=0,\\
\gamma F_{1}\left(  \alpha;\beta_{1},\beta_{2};\gamma,x,y\right)
+\gamma\left(  x-1\right)  F_{1}\left(  \alpha;\beta_{1}+1,\beta_{2}%
;\gamma,x,y\right)   & \nonumber\\
-\left(  \gamma-\alpha\right)  xF_{1}\left(  \alpha;\beta_{1}+1,\beta
_{2};\gamma+1,x,y\right)   &  =0,\label{c}\\
\gamma F_{1}\left(  \alpha;\beta_{1},\beta_{2};\gamma,x,y\right)
+\gamma\left(  y-1\right)  F_{1}\left(  \alpha;\beta_{1},\beta_{2}%
+1;\gamma,x,y\right)   & \nonumber\\
-\left(  \gamma-\alpha\right)  yF_{1}\left(  \alpha;\beta_{1},\beta
_{2}+1;\gamma+1,x,y\right)   &  =0. \label{d}%
\end{align}

It is straightforward to generalize the above relations and prove the
following $K+2$ recurrence relations for the $D$-type Lauricella functions:
\cite{1707.01281}
\begin{align}
\left(  \alpha-\underset{i}{\sum}\beta_{i}\right)  F_{D}^{(K)}\left(
\alpha;\beta_{1},...,\beta_{K};\gamma;x_{1},...,x_{K}\right)  -\alpha
F_{D}^{(K)}\left(  \alpha+1;\beta_{1},...,\beta_{K};\gamma;x_{1}%
,...,x_{K}\right)   & \nonumber\\
+\beta_{1}F_{D}^{(K)}\left(  \alpha;\beta_{1}+1,...,\beta_{K};\gamma
;x_{1},...,x_{K}\right)  +...+\beta_{K}F_{D}^{(K)}\left(  \alpha;\beta
_{1},...,\beta_{K}+1;\gamma;x_{1},...,x_{K}\right)   &  =0,\\
\gamma F_{D}^{(K)}\left(  \alpha;\beta_{1},...,\beta_{K};\gamma;x_{1}%
,...,x_{K}\right)  -\left(  \gamma-\alpha\right)  F_{D}^{(K)}\left(
\alpha;\beta_{1},...,\beta_{K};\gamma+1;x_{1},...,x_{K}\right)   & \nonumber\\
-\alpha F_{D}^{(K)}\left(  \alpha+1;\beta_{1},...,\beta_{K};\gamma
+1;x_{1},...,x_{K}\right)   &  =0,\\
\gamma F_{D}^{(K)}\left(  \alpha;\beta_{1},...,\beta_{m},...,\beta_{K}%
;\gamma;x_{1},...,x_{m},...,x_{K}\right)   & \nonumber\\
+\gamma(x_{m}-1)F_{D}^{(K)}\left(  \alpha;\beta_{1},...,\beta_{m}%
+1,...,\beta_{K};\gamma;x_{1},...,x_{m},...,x_{K}\right)   & \nonumber\\
+(\alpha-\gamma)x_{m}F_{D}^{(K)}\left(  \alpha;\beta_{1},...,\beta
_{m}+1,...,\beta_{K};\gamma+1;x_{1},...,x_{m},...,x_{K}\right)   &  =0,
\label{NN}%
\end{align}
where $m=1,2,...,K$. In the case of $K=2$, Equation (\ref{NN}) reduces to the Appell
recurrence relations in Equations (\ref{c}) and   (\ref{d}).

To simplify the notation, we omit those arguments of $F_{D}^{(K)}$ that
remain the same in the rest of the paper. Then, the above $K+2$ recurrence
relations can be expressed as%
\begin{align}
\left(  \alpha-\underset{i}{\sum}\beta_{i}\right)  F_{D}^{(K)}-\alpha
F_{D}^{(K)}\left(  \alpha+1\right)  +\beta_{1}F_{D}^{(K)}\left(  \beta
_{1}+1\right)  +...+\beta_{K}F_{D}^{(K)}\left(  \beta_{K}+1\right)   &
=0,\label{NNN1}\\
\gamma F_{D}^{(K)}-\left(  \gamma-\alpha\right)  F_{D}^{(K)}\left(
\gamma+1\right)  -\alpha F_{D}^{(K)}\left(  \alpha+1;\gamma+1\right)   &
=0,\label{NNN2}\\
\gamma F_{D}^{(K)}+\gamma(x_{m}-1)F_{D}^{(K)}\left(  \beta_{m}+1\right)
+(\alpha-\gamma)x_{m}F_{D}^{(K)}\left(  \beta_{m}+1,;\gamma+1\right)   &  =0.
\label{NNN}%
\end{align}

\begin{figure}[t]
\subfloat[]{
\includegraphics[width=.35\linewidth]{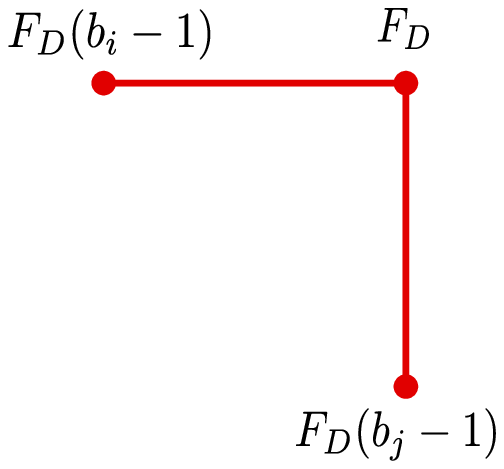} } \hspace{2cm}
\subfloat[]{
\includegraphics[width=.4\linewidth]{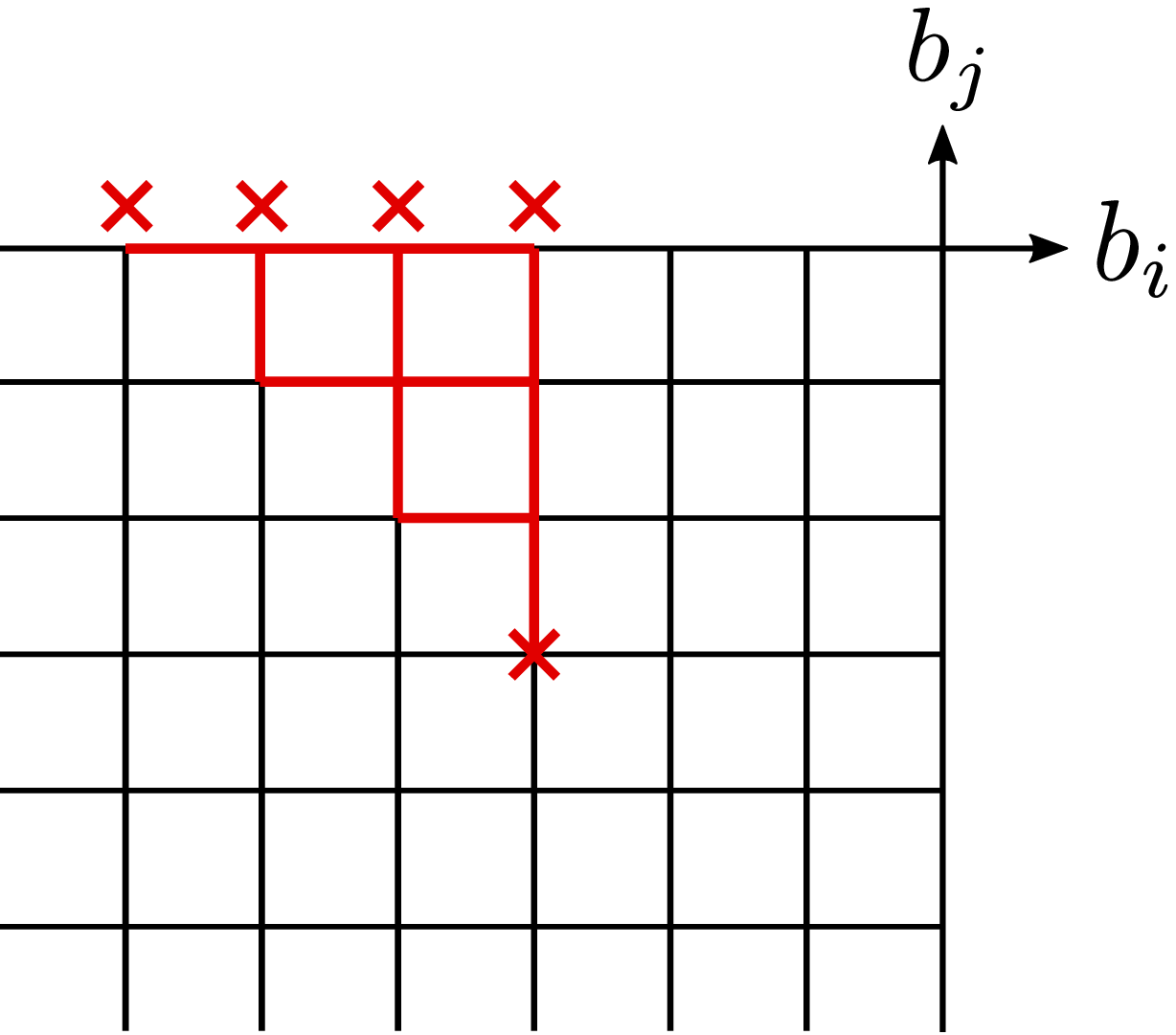} } \caption{(a) The three
neighborhood points are related by a recurrence relation. (b) The Lauricella fucntions can be reduced to the Gauss hypergeometry functions by decreasing their parameters $b_i$ to $0$ using the recurence relations.}
\label{pic}
\end{figure}

To proceed, we first consider the two recurrence relations from Equation (\ref{NNN})
for $m=i$, $j$ with $i\neq j$,%
\begin{equation}
cF_{D}^{(K)}+\gamma(x_{i}-1)F_{D}^{(K)}\left(  \beta_{i}+1\right)
+(\alpha-\gamma)x_{i}F_{D}^{(K)}\left(  \beta_{i}+1;\gamma+1\right)  =0,
\end{equation}%
\begin{equation}
\gamma F_{D}^{(K)}+\gamma(x_{j}-1)F_{D}^{(K)}\left(  \beta_{j}+1\right)
+(\alpha-\gamma)x_{j}F_{D}^{(K)}\left(  \beta_{j}+1;\gamma+1\right)  =0,
\end{equation}
By shifting $\beta_{i,j}$ to $\beta_{i,j}-1$ and combining the above two
equations to eliminate the $F_{D}^{(K)}\left(  c+1\right)  $ term, we obtain
the following key recurrence relation \cite{1707.01281}:
\begin{equation}
x_{j}F_{D}^{(K)}\left(  \beta_{i}-1\right)  -x_{i}F_{D}^{(K)}\left(  \beta
_{j}-1\right)  +\left(  x_{i}-x_{j}\right)  F_{D}^{(K)}=0. \label{key}%
\end{equation}
One can repeatedly apply Equation (\ref{key}) to the Lauricella functions in the LSSA
in\linebreak Equation (\ref{st1}) and end up with an expression that expresses $F_{D}%
^{(K)}(\beta_{1},\beta_{2},...\beta_{K})$ \linebreak  in terms of $F_{D}^{(K-1)}(\beta
_{1},..\beta_{i-1},\beta_{i+1}...\beta_{j}^{\prime},...\beta_{K})$, $\beta
_{j}^{\prime}=\beta_{j},\beta_{j}-1,...,\beta_{j}-\left\vert \beta
_{i}\right\vert $ or \linebreak $F_{D}^{(K-1)}(\beta_{1},...\beta_{i}^{\prime}%
,...\beta_{j-1},\beta_{j+1},...\beta_{K})$, $\beta_{i}^{\prime}=\beta_{i}%
,\beta_{i}-1,...,\beta_{i}-\left\vert \beta_{j}\right\vert $ (assume $i<j$). We
can repeat the above process to decrease the value of $K$ and reduce all the
Lauricella functions $F_{D}^{(K)}$ in the LSSA to the Gauss hypergeometry
functions $F_{D}^{(1)}=$ $_{2}F_{1}(\alpha,\beta,\gamma,x)$ as shown in
Figure \ref{pic}.

\subsection{Solving all the LSSAs}

In the last subsection, we expressed all the LSSAs in terms of the Gauss
hypergeometry functions $F_{D}^{(1)}=$ $_{2}F_{1}(\alpha,\beta,\gamma,x)$. In
this subsection, we further reduce the Gauss hypergeometry functions by
deriving a multiplication theorem for them and solve all the LSSAs in terms of
one single amplitude.

We begin with Taylor's theorem:
\begin{equation}
f(x+y)=\sum_{n=0}^{\infty}\frac{y^{n}}{n!}\frac{d^{n}}{dx^{n}}f(x).
\end{equation}
{By} replacing $y$ by $(y-1)x$, we get the identity
\begin{equation}
f(xy)=\sum_{n=0}^{\infty}\frac{(y-1)^{n}x^{n}}{n!}\frac{d^{n}}{dx^{n}}f(x).
\end{equation}

One can then use the derivative relation of the Gauss hypergeometry function%
\begin{equation}
\frac{d^{n}}{dx^{n}}\text{ }_{2}F_{1}(\alpha,\beta,\gamma,x)=\frac
{(\alpha)_{n}(\beta)_{n}}{(\gamma)_{n}}\text{ }_{2}F_{1}(\alpha+n,\beta
+n,\gamma+n,x),
\end{equation}
where $(\alpha)_{n}=\alpha\cdot\left(  \alpha+1\right)  \cdots\left(
\alpha+n-1\right)  $ is the Pochhammer symbol, to obtain the following
multiplication theorem:
\begin{equation}
_{2}F_{1}(\alpha,\beta,\gamma,xy)=\sum_{n=0}^{\left\vert \beta\right\vert
}\frac{(y-1)^{n}x^{n}}{n!}\frac{(\alpha)_{n}(\beta)_{n}}{(\gamma)_{n}}\text{
}_{2}F_{1}(\alpha+n,\beta+n,\gamma+n,x). \label{M}%
\end{equation}
It is important to note that the summation in the above equation is up to a
finite integer $\left\vert \beta\right\vert $ given that $\beta$ is a nonpositive
integer for the cases of LSSA.

In particular, if we take $x=1$ in Equation (\ref{M}), we get the following relation:
\begin{align}
_{2}F_{1}(\alpha,\beta,\gamma,y)  &  =\sum_{n=0}^{\left\vert \beta\right\vert
}\frac{(y-1)^{n}}{n!}\frac{(\alpha)_{n}(\beta)_{n}}{(\gamma)_{n}}\text{ }%
_{2}F_{1}(\alpha+n,\beta+n,\gamma+n,1)\nonumber\\
&  =\sum_{n=0}^{\left\vert \beta\right\vert }\frac{(y-1)^{n}}{n!}\frac
{(\alpha)_{n}(\beta)_{n}}{(\gamma)_{n}}\frac{(-)^{n}(\gamma)_{n}}%
{(\gamma-\alpha-\beta)_{n}}\text{ }_{2}F_{1}(\alpha,\beta,\gamma,1). \label{F}%
\end{align}
By using the following example of the $15$ Gauss contiguous relations%
\begin{equation}
\{\gamma-2\beta+(\beta-\alpha)x\}_{2}F_{1}+\beta(1-x)_{2}F_{1}(\beta
+1)+(\beta-\gamma)_{2}F_{1}(\beta-1)=0, \label{F4}%
\end{equation}
and setting $x=1$, which eliminates the second term of Equation (\ref{F4}), we can reduce the
argument $\beta$ in $_{2}F_{1}(\alpha,\beta,c,1)$ to $\beta=-1$ or $0$, which
corresponds to vector or tachyon amplitudes in the LSSA. This completes the
proof that all the LSSAs calculated in Equation (\ref{st1}) can be solved through
various recurrence relations of Lauricella functions. Moreover, all the LSSAs
can be expressed in terms of one single four tachyon amplitude.

\subsection{Examples of Solving LSSA}

For illustration, in this subsection, we calculate the Lauricella functions
which correspond to the LSSA for levels $K=1,2,3$.

For $K=1$, there are three type of LSSA $(\alpha=-\frac{t}{2}-1,\gamma=\frac
{u}{2}+2)$%
\begin{align}
(\alpha_{-1}^{T})^{p_{1}}\text{, }F_{D}^{(1)}(\alpha,-p_{1},\gamma
-p_{1},1)\text{, }N  &  =p_{1},\\
(\alpha_{-1}^{P})^{q_{1}}\text{, }F_{D}^{(1)}(\alpha,-q_{1},\gamma
-q_{1},\left[  \tilde{z}_{1}^{P}\right]  )\text{, }N  &  =q_{1},\\
(\alpha_{-1}^{L})^{r_{1}}\text{, }F_{D}^{(1)}(\alpha,-r_{1},\gamma
-r_{1},\left[  \tilde{z}_{1}^{L}\right]  )\text{, }N  &  =r_{1}.
\end{align}
For $K=2$, there are six type of LSSA $(\omega=-1)$%
\begin{align}
(\alpha_{-1}^{T})^{p_{1}}(\alpha_{-1}^{P})^{q_{1}}\text{, }F_{D}^{(2)}%
(\alpha,-p_{1},-q_{1},\gamma-p_{1}-q_{1},1,\left[  \tilde{z}_{1}^{P}\right]
)\text{,}N  &  =p_{1}+q_{1},\\
(\alpha_{-1}^{T})^{p_{1}}(\alpha_{-1}^{L})^{r_{1}}\text{, }F_{D}^{(2)}%
(\alpha,-p_{1},-r_{1},\gamma-p_{1}-r_{1},1,\left[  \tilde{z}_{1}^{L}\right]
)\text{,}N  &  =p_{1}+r_{1},\\
(\alpha_{-1}^{P})^{q_{1}}(\alpha_{-1}^{L})^{r_{1}}\text{, }F_{D}^{(2)}%
(\alpha,-q_{1},-r_{1},\gamma-q_{1}-r_{1},\left[  \tilde{z}_{1}^{P}\right]
,\left[  \tilde{z}_{1}^{L}\right]  )\text{,}N  &  =q_{1}+r_{1},\\
(\alpha_{-2}^{T})^{p_{2}}\text{, }F_{D}^{(2)}(\alpha,-p_{2},-p_{2}%
,\gamma-2p_{2},1,1)\text{, }N  &  =2p_{2},\\
(\alpha_{-2}^{P})^{q_{2}}\text{, }F_{D}^{(2)}(\alpha,-q_{2},-q_{2}%
,\gamma-2q_{2},1-Z_{2}^{P},1-\omega Z_{2}^{P})\text{, }N  &  =2q_{2},\\
(\alpha_{-2}^{L})^{r_{2}}\text{, }F_{D}^{(2)}(\alpha,-r_{2},-r_{2}%
,\gamma-2r_{2},1-Z_{2}^{L},1-\omega Z_{2}^{L})\text{, }N  &  =2r_{2}.
\end{align}
For $K=3$, there are 10 types of LSSA $(\omega_{1}=-1,\omega_{2}=\frac{\left(
-1+i\sqrt{3}\right)  /2}{2})$
\begin{align}
(\alpha_{-1}^{T})^{p_{1}}(\alpha_{-1}^{P})^{q_{1}}(\alpha_{-1}^{L})^{r_{1}%
}\text{, }F_{D}^{(3)}(\alpha,-p_{1},-q_{1},-r_{1},\gamma-p_{1}-q_{1}%
-r_{1},1,\left[  \tilde{z}_{1}^{P}\right]  ,\left[  \tilde{z}_{1}^{L}\right]
)\text{, }N  &  =p_{1}+q_{1}+r_{1},\\
(\alpha_{-2}^{T})^{p_{2}}(\alpha_{-1}^{P})^{q_{1}}\text{, }F_{D}^{(3)}%
(\alpha,-p_{2},-p_{2},-q_{1},\gamma-2p_{2}-q_{1},1,1,\left[  \tilde{z}_{1}%
^{P}\right]  )\text{, }N  &  =2p_{2}+q_{1},\\
(\alpha_{-2}^{T})^{p_{2}}(\alpha_{-1}^{L})^{r_{1}}\text{, }F_{D}^{(3)}%
(\alpha,-p_{2},-p_{2},-r_{1},\gamma-2p_{2}-r_{1},1,1,\left[  \tilde{z}_{1}%
^{L}\right]  )\text{, }N  &  =2p_{2}+r_{1},\\
(\alpha_{-1}^{T})^{p_{1}}(\alpha_{-2}^{P})^{q_{2}}\text{, }F_{D}^{(3)}%
(\alpha,-p_{1},-q_{2},-q_{2},\gamma-2q_{2}-p_{1},1,1-Z_{2}^{P},1-\omega
_{1}Z_{2}^{P})\text{, }N  &  =2q_{2}+p_{1},\\
(\alpha_{-2}^{P})^{q_{2}}(\alpha_{-1}^{L})^{r_{1}}\text{, }F_{D}^{(3)}%
(\alpha,-q_{2},-q_{2},-r_{1},\gamma-2q_{2}-r_{1},1-Z_{2}^{P},1-\omega_{1}%
Z_{2}^{P},\left[  \tilde{z}_{1}^{L}\right]  )\text{, }N  &  =2q_{2}+r_{1},\\
(\alpha_{-1}^{T})^{p_{1}}(\alpha_{-2}^{L})^{r_{2}}\text{, }F_{D}^{(3)}%
(\alpha,,-p_{1},-r_{2},-r_{2},\gamma-2r_{2}-p_{1},1,1-Z_{2}^{L},1-\omega
_{1}Z_{2}^{L})\text{, }N  &  =2r_{2}+p_{1}.\\
(\alpha_{-1}^{P})^{q_{1}}(\alpha_{-2}^{L})^{r_{2}}\text{, }F_{D}^{(3)}%
(\alpha,,-q_{1},-r_{2},-r_{2},\gamma-2r_{2}-q_{1},\left[  \tilde{z}_{1}%
^{P}\right]  ,1-Z_{2}^{L},1-\omega_{1}Z_{2}^{L})\text{, }N  &  =2r_{2}%
+q_{1}.\\
(\alpha_{-3}^{T})^{p_{3}}\text{, }F_{D}^{(3)}(\alpha,-p_{3},-p_{3}%
,-p_{3},\gamma-3p_{3},1,1,1)\text{, }N  &  =3p_{3},\\
(\alpha_{-3}^{P})^{q_{3}}\text{, }F_{D}^{(3)}(\alpha,-q_{3},-q_{3}%
,-q_{3},\gamma-3q_{3},1-Z_{3}^{P},1-\omega_{2}Z_{3}^{P},1-\omega_{2}^{2}%
Z_{3}^{P})\text{, }N  &  =3q_{3},\\
(\alpha_{-3}^{L})^{r_{3}}\text{, }F_{D}^{(3)}(\alpha,-r_{3},-r_{3}%
,-r_{3},\gamma-3r_{3},1-Z_{3}^{L},1-\omega_{2}Z_{3}^{L},1-\omega_{2}^{2}%
Z_{3}^{L})\text{, }N  &  =3r_{3}.
\end{align}
All the LSSAs for $K=2,3$ can be reduced through the recurrence relations in \linebreak
\mbox{Equation (\ref{key})} and expressed in terms of those of $K=1.$ Furthermore, all
resulting LSSAs for $K=1$ can be further reduced by applying Equations (\ref{F}) and (\ref{F4}) and finally expressed in terms of one single LSSA.

\subsection{SL($K+3$,C) Symmetry and Recurrence Relations}

In this subsection, we use the recurrence relations of the
$D$-type \linebreak $F_{D}^{(K)}\left(  \alpha;\beta_{1},...,\beta_{K};\gamma
;x_{1},...,x_{K}\right)  $ to reproduce the Cartan subalgebra and simple root
system of $SL(K+3,%
\mathbb{C}
)$ with rank $K+2$. We first review the case of the $SL(4,%
\mathbb{C}
)$ symmetry group, and then extend it to the general case of $SL(K+3,%
\mathbb{C}
)$ Symmetry.

\subsubsection{$SL(4,%
\mathbb{C}
)$ Symmetry}

We first relate the $SL(4,%
\mathbb{C}
)$ group to the recurrence relations of $F_{D}^{\left(  1\right)  }\left(
\alpha;\beta;\gamma;x\right)  $ or of the LSSA in Equation (\ref{rel}). For our
purpose, there are $K+2=1+2=3$ recurrence relations among $F_{D}^{\left(
1\right)  }\left(  \alpha;\beta;\gamma;x\right)  $ or Gauss hypergeometry functions%

\begin{align}
\left(  \alpha-\beta\right)  F_{D}^{\left(  1\right)  }-\alpha F_{D}^{\left(
1\right)  }\left(  \alpha+1\right)  +\beta F_{D}^{\left(  1\right)  }\left(
\beta+1\right)   &  =0,\label{R1}\\
\gamma F_{D}^{\left(  1\right)  }-\left(  \gamma-\alpha\right)  F_{D}^{\left(
1\right)  }\left(  \gamma+1\right)  -\alpha F_{D}^{\left(  1\right)  }\left(
\alpha+1;\gamma+1\right)   &  =0,\label{R2}\\
\gamma F_{D}^{\left(  1\right)  }+\gamma\left(  x-1\right)  F_{D}^{\left(
1\right)  }\left(  \beta+1\right)  -\left(  \gamma-\alpha\right)
xF_{D}^{\left(  1\right)  }\left(  \beta+1;\gamma+1\right)   &  =0, \label{R3}%
\end{align}
which can be used to reproduce the Cartan subalgebra and simple root system of
the $SL(4,%
\mathbb{C}
)$ group with rank $3$.

With the identification in Equation (\ref{id}), the first recurrence relation in
Equation (\ref{R1}) can be rewritten~as
\begin{equation}
\frac{\left(  \alpha-\beta\right)  f_{ac}^{b}\left(  \alpha;\beta
;\gamma;x\right)  }{B\left(  \gamma-\alpha,\alpha\right)  a^{\alpha}b^{\beta
}c^{\gamma}}-\frac{\alpha f_{ac}^{b}\left(  \alpha+1;\beta;\gamma;x\right)
}{B\left(  \gamma-\alpha-1,\alpha+1\right)  a^{\alpha+1}b^{\beta}c^{\gamma}%
}+\frac{\beta f_{ac}^{b}\left(  \alpha;\beta+1;\gamma;x\right)  }{B\left(
\gamma-\alpha,\alpha\right)  a^{\alpha}b^{\beta+1}c^{\gamma}}=0.
\end{equation}
By using the identity%
\begin{equation}
B\left(  \gamma-\alpha-1,\alpha+1\right)  =\frac{\Gamma\left(  \gamma
-\alpha-1\right)  \Gamma\left(  \alpha+1\right)  }{\Gamma\left(
\gamma\right)  }=\frac{\alpha}{\gamma-\alpha-1}\frac{\Gamma\left(
\gamma-\alpha\right)  \Gamma\left(  \alpha\right)  }{\Gamma\left(
\gamma\right)  },
\end{equation}
the recurrence relation then becomes%
\begin{equation}
\left(  \alpha-\beta\right)  f_{ac}^{b}\left(  \alpha;\beta;\gamma;x\right)
-\frac{\gamma-\alpha-1}{a}f_{ac}^{b}\left(  \alpha+1;\beta;\gamma;x\right)
+\frac{\beta}{b}f_{ac}^{b}\left(  \alpha;\beta+1;\gamma;x\right)  =0,
\end{equation}
or%
\begin{equation}
\left(  \alpha-\beta-\frac{E_{\alpha}}{a}+\frac{E_{\beta}}{b}\right)
f_{ac}^{b}\left(  \alpha;\beta;\gamma;x\right)  =0,
\end{equation}
which means%
\begin{equation}
\left[  \alpha-\beta-\left(  x\partial_{x}+a\partial_{a}\right)  +\left(
x\partial_{x}+b\partial_{b}\right)  \right]  f_{ac}^{b}\left(  \alpha
;\beta;\gamma;x\right)  =0,
\end{equation}
or%
\begin{equation}
\left[  \left(  \alpha-J_{\alpha}\right)  -\left(  \beta-J_{\beta}\right)
\right]  f_{ac}^{b}\left(  \alpha;\beta;\gamma;x\right)  =0. \label{R11}%
\end{equation}

Similarly, for the second recurrence relation in Equation (\ref{R2}), we obtain%
\begin{equation}
\left[  c\left(  \gamma-\beta\right)  -E_{\gamma}+\frac{E_{\alpha\gamma}}%
{a}\right]  f_{ac}^{b}\left(  \alpha;\beta;\gamma;x\right)  =0.
\end{equation}
which means%
\begin{equation}
\left[  \left(  \gamma-c\partial_{c}\right)  -\left(  \beta-b\partial
_{b}\right)  \right]  f_{ac}^{b}\left(  \alpha;\beta;\gamma;x\right)  =0,
\end{equation}
or%
\begin{equation}
\left[  \left(  \gamma-J_{\gamma}\right)  -\left(  \beta-J_{\beta}\right)
\right]  f_{ac}^{b}\left(  \alpha;\beta;\gamma;x\right)  =0. \label{R22}%
\end{equation}

Finally, the third recurrence relation in Equation (\ref{R3}) can be rewritten as
\begin{equation}
\left[  b\beta+\left(  x-1\right)  E_{\beta}-\frac{xE_{\beta\gamma}}%
{c}\right]  f_{ac}^{b}\left(  \alpha;\beta;\gamma;x\right)  =0,
\end{equation}
which gives after some computation%
\begin{equation}
\left(  \beta-J_{\beta}\right)  f_{ac}^{b}\left(  \alpha;\beta;\gamma
;x\right)  =0. \label{R33}%
\end{equation}

It is easy to see that Equations (\ref{R11}),  (\ref{R22}) and (\ref{R33}) imply
the last three equations of Equation (\ref{op}) or the Cartan subalgebra in
Equation (\ref{cartan}), as expected.

In addition to the Cartan subalgebra, we need to derive the operations of the
$\{E_{\alpha},E_{\beta},E_{\gamma}\}$ from the recurrence relations. With the
operations of Cartan subalgebra and $\{E_{\alpha},E_{\beta},E_{\gamma}\}$, one
can reproduce the entirety of $SL(4,%
\mathbb{C}
)$ algebra.

We first use the operation of $E_{\alpha,\beta}$ in Equation (\ref{op}) to express
Equation (\ref{R1}) in the following two~ways:
\begin{align}
\left(  \alpha-\beta-\frac{E_{a}}{a}\right)  f_{ac}^{b}\left(  \alpha
;\beta;\gamma;x\right)  +\frac{\beta}{b}f_{ac}^{b}\left(  \alpha
;\beta+1;\gamma;x\right)   &  =0,\\
\left(  \alpha-\beta+\frac{E_{\beta}}{b}\right)  f_{ac}^{b}\left(
\alpha;\beta;\gamma;x\right)  -\frac{\left(  \gamma-\alpha-1\right)  }%
{a}f_{ac}^{b}\left(  \alpha+1;\beta;\gamma;x\right)   &  =0,
\end{align}
which, by using the definition of $E_{\alpha,\beta}$ in Equation (\ref{def}), become%
\begin{align}
\left(  \alpha-\beta-\frac{a\left(  x\partial_{x}+a\partial_{a}\right)  }%
{a}\right)  f_{ac}^{b}\left(  \alpha;\beta;\gamma;x\right)   &  =-\frac{\beta
f_{ac}^{b}\left(  \alpha;\beta+1;\gamma;x\right)  }{b},\\
\left(  \alpha-\beta+\frac{b\left(  x\partial_{x}+b\partial_{b}\right)  }%
{b}\right)  f_{ac}^{b}\left(  \alpha;\beta;\gamma;x\right)   &  =\frac{\left(
\gamma-\alpha-1\right)  f_{ac}^{b}\left(  \alpha+1;\beta;\gamma;x\right)  }%
{a},
\end{align}
which in turn imply%
\begin{align}
\left[  b\left(  b\partial_{b}+x\partial_{x}\right)  \right]  f_{ac}%
^{b}\left(  \alpha;\beta;\gamma;x\right)   &  =E_{\beta}f_{ac}^{b}\left(
\alpha;\beta;\gamma;x\right)  =\beta f_{ac}^{b}\left(  \alpha;\beta
+1;\gamma;x\right)  ,\label{ok1}\\
\left[  a\left(  a\partial_{a}+x\partial_{x}\right)  \right]  f_{ac}%
^{b}\left(  \alpha;\beta;\gamma;x\right)   &  =E_{\alpha}f_{ac}^{b}\left(
\alpha;\beta;\gamma;x\right)  =\left(  \gamma-\alpha-1\right)  f_{ac}%
^{b}\left(  \alpha+1;\beta;\gamma;x\right)  , \label{ok2}%
\end{align}

The above Equations (\ref{ok1}) and (\ref{ok2}) are consistent with the operation
of $E_{\alpha,\beta}$ in\linebreak Equation (\ref{op}).

Finally, we check the operation of $E_{\gamma}$. Note that Equation (\ref{R2}) can be
written as%
\begin{equation}
\frac{\gamma f_{ac}^{b}\left(  \alpha;\beta;\gamma;x\right)  }{B\left(
\gamma-\alpha,\alpha\right)  a^{\alpha}b^{\beta}c^{\gamma}}-\frac{\left(
\gamma-\alpha\right)  f_{ac}^{b}\left(  \alpha;\beta;\gamma+1;x\right)
}{\frac{\left(  \gamma-\alpha\right)  }{\gamma}B\left(  \gamma-\alpha
,\alpha\right)  a^{\alpha}b^{\beta}c^{\gamma+1}}-\frac{\alpha f_{ac}%
^{b}\left(  \alpha+1;\beta;\gamma+1;x\right)  }{\frac{\alpha}{\gamma}B\left(
\gamma-\alpha,\alpha\right)  a^{\alpha+1}b^{\beta}c^{\gamma+1}}=0,
\end{equation}
which gives%
\begin{equation}
f_{ac}^{b}\left(  \alpha;\beta;\gamma;x\right)  -\frac{1}{c}f_{ac}^{b}\left(
\alpha;\beta;\gamma+1;x\right)  -\frac{1}{ac}f_{ac}^{b}\left(  \alpha
+1;\beta;\gamma+1;x\right)  =0.
\end{equation}

Using the definition and operation of $E_{\alpha\gamma}$ in Equation (\ref{def}), we
obtain%
\begin{equation}
f_{ac}^{b}\left(  \alpha;\beta;\gamma;x\right)  -\frac{1}{c}f_{ac}^{b}\left(
\alpha;\beta;\gamma+1;x\right)  -\frac{E_{\alpha\gamma}}{ac\left(
\beta-\gamma\right)  }f_{ac}^{b}\left(  \alpha;\beta;\gamma;x\right)
=0,\nonumber
\end{equation}
which gives%
\begin{equation}
f_{ac}^{b}\left(  \alpha;\beta;\gamma;x\right)  -\frac{ac\left[  \left(
1-x\right)  \partial_{x}-a\partial_{a}\right]  f_{ac}^{b}\left(  \alpha
;\beta;\gamma;x\right)  }{ac\left(  \beta-\gamma\right)  }=\frac{f_{ac}%
^{b}\left(  \alpha;\beta;\gamma+1;x\right)  }{c}.
\end{equation}
After some simple computation, we get
\begin{equation}
-c\left[  b\partial_{b}-c\partial_{c}-\left(  1-x\right)  \partial
_{x}+a\partial_{a}\right]  f_{ac}^{b}\left(  \alpha;\beta;\gamma;x\right)
=E_{\gamma}f_{ac}^{b}\left(  \alpha;\beta;\gamma;x\right)  =\left(
\gamma-\beta\right)  f_{ac}^{b}\left(  \alpha;\beta;\gamma+1;x\right)
,\nonumber
\end{equation}
which is consistent with the operation of $E_{\gamma}$ in Equation (\ref{op}).

Thus, we have shown that the extended LSSAs $f_{ac}^{b}\left(
\alpha;\beta;\gamma;x\right)  $ in Equation (\ref{id}) with arbitrary $a$ and $c$
form an infinite-dimensional representation of the $SL(4,%
\mathbb{C}
)$ group. Moreover, the $3$ recurrence relations among the LSSAs can be used to
reproduce the Cartan subalgebra and simple root system of the $SL(4,%
\mathbb{C}
)$ group with rank $3$. The recurrence relations are thus equivalent to the
representation of the $SL(4,%
\mathbb{C}
)$ symmetry group.

\subsubsection{$SL(K+3,%
\mathbb{C}
)$ Symmetry}

The $K+2$ fundamental recurrence relations among $F_{D}^{\left(  K\right)
}\left(  \alpha;\beta;\gamma;x\right)  $ or the Lauricella functions are
listed in Equations (\ref{NNN1})--(\ref{NNN}). In the following, we show that the
three types of recurrence relations above imply the Cartan subalgebra of the
$SL(K+3,%
\mathbb{C}
)$ group with rank $K+2$.

With the identification in Equation (\ref{id2}), the first type of recurrence
relation in\linebreak  Equation (\ref{NNN1}) can be rewritten as%
\begin{equation}
\left(  \alpha-\underset{j}{\sum}\beta_{j}\right)  f_{ac}^{b_{1}\cdots b_{K}%
}-\frac{E^{\alpha}f_{ac}^{b_{1}\cdots b_{K}}\left(  \alpha\right)  }%
{a}+\underset{j}{%
{\displaystyle\sum}
}\frac{E^{\beta_{j}}f_{ac}^{b_{1}\cdots b_{K}}\left(  \beta_{j}\right)
}{b_{j}}=0,
\end{equation}
which gives%
\begin{equation}
\left(  \alpha-\underset{j}{\sum}\beta_{j}\right)  f_{ac}^{b_{1}\cdots b_{K}%
}-\left(  \underset{j}{%
{\displaystyle\sum}
}x_{j}\partial_{j}+a\partial_{a}\right)  f_{ac}^{b_{1}\cdots b_{K}%
}+\underset{j}{%
{\displaystyle\sum}
}\left(  x_{j}\partial_{j}+b_{j}\partial_{b_{j}}\right)  f_{ac}^{b_{1}\cdots
b_{K}}=0
\end{equation}
or%
\begin{equation}
\left[  \left(  \alpha-a\partial_{a}\right)  +\underset{j}{\sum}\left(
\beta_{j}-b_{j}\partial_{b_{j}}\right)  \right]  f_{ac}^{b_{1}\cdots b_{K}}=0,
\label{car1}%
\end{equation}
which means%
\begin{equation}
\left[  \left(  \alpha-J_{\alpha}\right)  +\underset{j}{\sum}\left(  \beta
_{j}-J_{\beta_{j}}\right)  \right]  f_{ac}^{b_{1}\cdots b_{K}}=0.
\label{car11}%
\end{equation}

The second type of recurrence relation in Equation (\ref{NNN2}) can be rewritten as%
\begin{equation}
f_{ac}^{b_{1}\cdots b_{K}}-\frac{E^{\gamma}f_{ac}^{b_{1}\cdots b_{K}}\left(
\gamma\right)  }{c\left(  \gamma-\underset{j}{%
{\displaystyle\sum}
}\beta_{j}\right)  }-\frac{E^{\alpha\gamma}f_{ac}^{b_{1}\cdots b_{K}}\left(
\alpha;\gamma\right)  }{ac\left(  \underset{j}{%
{\displaystyle\sum}
}\beta_{j}-\gamma\right)  }=0,
\end{equation}
which gives
\begin{equation}
\left[  \gamma-\underset{j}{%
{\displaystyle\sum}
}\beta_{j}-\left(  \underset{j}{%
{\displaystyle\sum}
}\left(  1-x_{j}\right)  \partial_{x_{j}}+c\partial_{c}-a\partial
_{a}-\underset{j}{%
{\displaystyle\sum}
}b_{j}\partial_{b_{j}}\right)  +\left(  \underset{j}{%
{\displaystyle\sum}
}\left(  1-x_{j}\right)  \partial_{x_{j}}-a\partial_{a}\right)  \right]
f_{ac}^{b_{1}\cdots b_{K}}=0
\end{equation}
or%
\begin{equation}
\left[  \left(  \gamma-c\partial_{c}\right)  -\underset{j}{%
{\displaystyle\sum}
}\left(  \beta_{j}-b_{j}\partial_{b_{j}}\right)  \right]  f_{ac}^{b_{1}\cdots
b_{K}}=0. \label{car2}%
\end{equation}
Equation (\ref{car2}) can be written as%
\begin{equation}
\left[  \left(  \gamma-J_{\gamma}\right)  -\underset{j}{%
{\displaystyle\sum}
}\left(  \beta_{j}-J_{\beta_{j}}\right)  \right]  f_{ac}^{b_{1}\cdots b_{K}%
}=0. \label{car22}%
\end{equation}

The third type of recurrence relation in Equation (\ref{NNN}) can be rewritten as
($m=1,2,...K$)%
\begin{equation}
f_{ac}^{b_{1}\cdots b_{K}}+\frac{(x_{m}-1)E^{\beta_{m}}f_{ac}^{b_{1}\cdots
b_{K}}}{b_{m}\beta_{m}}-\frac{x_{m}E^{\beta_{m}\gamma}f_{ac}^{b_{1}\cdots
b_{K}}}{b_{m}c\beta_{m}}=0,
\end{equation}
which gives%
\begin{equation}
\beta_{m}f_{ac}^{b_{1}\cdots b_{K}}+(x_{m}-1)\left(  x_{m}\partial_{m}%
+b_{m}\partial_{b_{m}}\right)  f_{ac}^{b_{1}\cdots b_{K}}-x_{m}\left[  \left(
x_{m}-1\right)  \partial_{x_{m}}+b_{m}\partial_{b_{m}}\right]  f_{ac}%
^{b_{1}\cdots b_{K}}=0
\end{equation}
or%
\begin{equation}
\left(  \beta_{m}-b_{m}\partial_{b_{m}}\right)  f_{ac}^{b_{1}\cdots b_{K}}=0.
\label{car3}%
\end{equation}
In the above calculation, we have used the definition and operation of
$E^{\beta_{m}\gamma}$ in\linebreak Equation (\ref{def2}) and Equation (\ref{op2}), respectively.

Equation (\ref{car3}) can be written as%
\begin{equation}
\left(  \beta_{m}-J_{\beta_{m}}\right)  f_{ac}^{b_{1}\cdots b_{K}%
}=0,m=1,2,...K. \label{car33}%
\end{equation}

It is important to see that Equations (\ref{car11}),  (\ref{car22}) and
 (\ref{car33}) imply the last three equations of Equation (\ref{op2}) or the
Cartan subalgebra of $SL(K+3,%
\mathbb{C}
)$ as expected.

In addition to the Cartan subalgebra, we need to derive the operations of the \linebreak
$\{E^{\alpha},E^{\beta_{k}},E^{\gamma}\}$ from the recurrence relations. With
the operations of Cartan subalgebra and $\{E^{\alpha},E^{\beta_{k}},E^{\gamma
}\}$, one can reproduce the whole $SL(K+3,%
\mathbb{C}
)$ algebra. The calculations of $E^{\alpha}$ and $E^{\gamma}$ are
straightforward and are similar to the case of $SL(4,%
\mathbb{C}
)$ in the previous section. Here, we present only the calculation of
$E^{\beta_{k}}$. The recurrence relation in \linebreak Equation (\ref{NNN1}) can be rewritten
as%
\begin{equation}
\left(  \alpha-\underset{j}{\sum}\beta_{j}\right)  f_{ac}^{b_{1}\cdots b_{K}%
}-\frac{E^{\alpha}f_{ac}^{b_{1}\cdots b_{K}}\left(  \alpha\right)  }%
{a}+\underset{j\neq k}{%
{\displaystyle\sum}
}\frac{E^{\beta_{j}}f_{ac}^{b_{1}\cdots b_{K}}\left(  \beta_{j}\right)
}{b_{j}}+\frac{\beta_{k}f_{ac}^{b_{1}\cdots b_{K}}\left(  \beta_{k}+1\right)
}{b_{k}}=0.
\end{equation}
After the operation of $E^{\beta_{j}}$, we obtain
\begin{equation}
\left(  \alpha-\underset{j}{\sum}\beta_{j}\right)  f_{ac}^{b_{1}\cdots b_{K}%
}-\left(  \underset{j}{%
{\displaystyle\sum}
}x_{j}\partial_{j}+a\partial_{a}\right)  f_{ac}^{b_{1}\cdots b_{K}%
}+\underset{j\neq k}{%
{\displaystyle\sum}
}\left(  x_{j}\partial_{j}+b_{j}\partial_{b_{j}}\right)  f_{ac}^{b_{1}\cdots
b_{K}}=\frac{-\beta_{k}f_{ac}^{b_{1}\cdots b_{K}}\left(  \beta_{k}+1\right)
}{b_{k}},\nonumber
\end{equation}
which gives the consistent result%
\begin{equation}
b_{k}\left(  b_{k}\partial_{b_{k}}+x_{k}\partial_{k}\right)  f_{ac}%
^{b_{1}\cdots b_{K}}\left(  \beta_{k}\right)  =E^{\beta_{k}}f_{ac}%
^{b_{1}\cdots b_{K}}=\beta_{k}f_{ac}^{b_{1}\cdots b_{K}}\left(  \beta
_{k}+1\right)  ,k=1,2,...K.
\end{equation}
In the above calculation, we have used the definitions and operations of
$E^{\beta_{k}}$ and $E^{\alpha}$ in Equation (\ref{def2}) and Equation (\ref{op2}), respectively.

The $K+2$ equations in Equations (\ref{car11}),  (\ref{car22}) and (\ref{car33})
together with $K+2$ equations for the operations $\{E^{\alpha},E^{\beta_{k}%
},E^{\gamma}\}$ are equivalent to the Cartan subalgebra and the simple root
system of $SL(K+3,%
\mathbb{C}
)$ with rank $K+2.$ With the Cartan subalgebra and the simple roots, one can
easily write the whole Lie algebra of the $SL(K+3,%
\mathbb{C}
)$ group. Thus, one can construct the Lie algebra from the recurrence relations
and vice versa.

In the previous subsections, it was shown that \cite{1707.01281} the $K+2$
recurrence relations among $F_{D}^{(K)}$ can be used to derive recurrence
relations among LSSAs and reduce the number of independent\ LSSAs from $\infty$
down to $1$. We conclude that the $SL(K+3,%
\mathbb{C}
)$ group can be used to derive an infinite number of recurrence relations among
LSSAs, and one can solve all the LSSA sand express them in terms of one amplitude.

\subsection{Lauricella Zero Norm States and Ward Identities}

In addition to the recurrence relations among LSSAs, there are on-shell stringy
Ward identities among LSSAs. These Ward identities can be derived from the
decoupling of two types of zero norm states (ZNS) in the old covariant first
quantized string spectrum. However, we show below that these Lauricella
zero norm states (LZNS) or the corresponding Lauricella Ward identities are
not good enough to solve all the LSSAs and express them in terms of
one amplitude.

On the other hand, in the last section, we have shown that by using (A)
recurrence relations of the LSSAs, (B) the multiplication theorem of the Gauss
hypergeometry function and (C) the explicit calculation of four tachyon
amplitudes, one can explicitly solve and calculate all LSSAs. This means that
the solvability of LSSAs through the calculations of (A), (B) and (C) implies the
validity of Ward identities. Ward identities cannot be independent
of the recurrence relations used in the last section; otherwise, there will be a
contradiction with the solvability of LSSAs.

In this section, we study some examples of Ward identities of LSSAs from
this point of view. Incidentally, high-energy zero norm states (HZNS)
\cite{ChanLee1,ChanLee2,CHL,CHLTY2, CHLTY1,susy} and the corresponding stringy
Ward identities at the fixed angle regime, Regge zero norm states (RZNS)
\cite{LY,AppellLY} and the corresponding Regge Ward identities at the Regge
regime have been studied previously. In particular, HZNS at the fixed angle
regime can be used to solve all the high energy SSAs
\cite{ChanLee1,ChanLee2,CHL,CHLTY2, CHLTY1,susy}.

\subsubsection{The Lauricella Zero Norm States}

We consider the set of Ward identities of the LSSA with three tachyons
and one arbitrary string state. Thus, we only need to consider polarizations
of the tensor states on the scattering plane since the amplitudes with
polarizations orthogonal to the scattering plane vanish.

There are two types of zero norm states (ZNS) in the old covariant first
quantum string spectrum:
\begin{align}
\text{Type I}  &  :L_{-1}\left\vert x\right\rangle ,\text{ where }%
L_{1}\left\vert x\right\rangle =L_{2}\left\vert x\right\rangle =0,\text{
}L_{0}\left\vert x\right\rangle =0;\label{ZN1}\\
\text{Type II}  &  :\left(  L_{-2}+\frac{3}{2}L_{-1}^{2}\right)  \left\vert
\tilde{x}\right\rangle ,\text{ where }L_{1}\left\vert \tilde{x}\right\rangle
=L_{2}\left\vert \tilde{x}\right\rangle =0,\text{ }(L_{0}+1)\left\vert
\tilde{x}\right\rangle =0.
\end{align}
While type I ZNS exists at any spacetime dimension, type II ZNS only
exists at $D=26$.

We begin with the case of mass level $M^{2}=2$. There is a type II ZNS%
\begin{equation}
\left[  \frac{1}{2}\alpha_{-1}\cdot\alpha_{-1}+\frac{5}{2}k\cdot\alpha
_{-2}+\frac{3}{2}(k\cdot\alpha_{-1})^{2}\right]  \left\vert 0,k\right\rangle ,
\label{2.1}%
\end{equation}
and a type I ZNS%

\begin{equation}
\lbrack\theta\cdot\alpha_{-2}+(k\cdot\alpha_{-1})(\theta\cdot\alpha
_{-1})]\left\vert 0,k\right\rangle ,\theta\cdot k=0. \label{2.2}%
\end{equation}
The three polarizations defined in Equations (\ref{aa})--(\ref{cc}) of the second
tensor state with momentum $k_{2}$ on the scattering plane satisfy the
completeness relation%
\begin{equation}
\eta_{\mu\nu}=\sum_{\alpha,\beta}e_{\mu}^{\alpha}e_{\nu}^{\beta}\eta
_{\alpha\beta}=diag(-1,1,1)
\end{equation}
where $\mu,\nu=0,1,2$ and $\alpha,\beta=P,L,T$. and $\alpha_{-1}^{T}=\sum
_{\mu}e_{\mu}^{T}\alpha_{-1}^{\mu}$, $\alpha_{-1}^{T}\alpha_{-2}^{L}=\sum
_{\mu,\nu}e_{\mu}^{T}e_{\nu}^{L}\alpha_{-1}^{\mu}\alpha_{-2}^{\nu}$ etc.

The type II ZNS in Equation (\ref{2.1}) gives the LZNS%
\begin{equation}
\left(  \sqrt{2}\alpha_{-2}^{P}+\alpha_{-1}^{P}\alpha_{-1}^{P}+\frac{1}%
{5}\alpha_{-1}^{L}\alpha_{-1}^{L}+\frac{1}{5}\alpha_{-1}^{T}\alpha_{-1}%
^{T}\right)  |0,k\rangle. \label{R2.3}%
\end{equation}

The type I ZNS in Equation (\ref{2.2}) gives two LZNSs:
\begin{equation}
(\alpha_{-2}^{T}+\sqrt{2}\alpha_{-1}^{P}\alpha_{-1}^{T})|0,k\rangle,
\label{R2.1}%
\end{equation}%
\begin{equation}
(\alpha_{-2}^{L}+\sqrt{2}\alpha_{-1}^{P}\alpha_{-1}^{L})|0,k\rangle.
\label{R2.2}%
\end{equation}
where $\alpha_{-1}^{T}=\sum_{\mu}e_{\mu}^{T}\alpha_{-1}^{\mu}$, $\alpha
_{-1}^{T}\alpha_{-2}^{L}=\sum_{\mu,\nu}e_{\mu}^{T}e_{\nu}^{L}\alpha_{-1}^{\mu
}\alpha_{-2}^{\nu}$ etc. The LZNSs in Equations (\ref{R2.1}) and  (\ref{R2.2})
correspond to choosing $\theta^{\mu}=e^{T}$ and $\theta^{\mu}=e^{L}$,
respectively. In conclusion, there are $3$ LZNSs at the mass level $M^{2}=$ $2$.

At the second massive level $M^{2}=4,$ there is a type I scalar ZNS,
\begin{equation}
\left[  \frac{17}{4}(k\cdot\alpha_{-1})^{3}+\frac{9}{2}(k\cdot\alpha
_{-1})(\alpha_{-1}\cdot\alpha_{-1})+9(\alpha_{-1}\cdot\alpha_{-2}%
)+21(k\cdot\alpha_{-1})(k\cdot\alpha_{-2})+25(k\cdot\alpha_{-3})\right]
\left\vert 0,k\right\rangle , \label{41}%
\end{equation}
a symmetric type I spin two ZNS,
\begin{equation}
\lbrack2\theta_{\mu\nu}\alpha_{-1}^{(\mu}\alpha_{-2}^{\nu)}+k_{\lambda}%
\theta_{\mu\nu}\alpha_{-1}^{\lambda\mu\nu}]\left\vert 0,k\right\rangle
,k\cdot\theta=\eta^{\mu\nu}\theta_{\mu\nu}=0,\theta_{\mu\nu}=\theta_{\nu\mu},
\label{42}%
\end{equation}
where $\alpha_{-1}^{\lambda\mu\nu}\equiv\alpha_{-1}^{\lambda}\alpha_{-1}^{\mu
}\alpha_{-1}^{\nu}$, and two vector ZNSs,
\begin{align}
\left[  \left(  \frac{5}{2}k_{\mu}k_{\nu}\theta_{\lambda}^{\prime}+\eta
_{\mu\nu}\theta_{\lambda}^{\prime}\right)  \mathcal{\alpha}_{-1}^{(\mu
\nu\lambda)}+9k_{\mu}\theta_{\nu}^{\prime}\mathcal{\alpha}_{-1}^{(\mu\nu
)}+6\theta_{\mu}^{\prime}\mathcal{\alpha}_{-1}^{\mu}\right]  \left\vert
0,k\right\rangle ,\theta\cdot k  &  =0,\label{43}\\
\left[  \left(  \frac{1}{2}k_{\mu}k_{\nu}\theta_{\lambda}+2\eta_{\mu\nu}%
\theta_{\lambda}\right)  \mathcal{\alpha}_{-1}^{(\mu\nu\lambda)}+9k_{\mu
}\theta_{\nu}\mathcal{\alpha}_{-1}^{[\mu\nu]}-6\theta_{\mu}\mathcal{\alpha
}_{-1}^{\mu}\right]  \left\vert 0,k\right\rangle ,\theta\cdot k  &  =0.
\label{44}%
\end{align}
Note that Equations (\ref{43}) and (\ref{44}) are linear combinations of a type I
and a type II ZNS. This completes the four ZNSs at the second massive level
$M^{2}=$ $4$.

The scalar ZNS in Equation (\ref{41}) gives the LZNS
\begin{equation}
\left[  25(\alpha_{-1}^{P})^{3}+9\alpha_{-1}^{P}(\alpha_{-1}^{L})^{2}%
+9\alpha_{-1}^{P}(\alpha_{-1}^{T})^{2}+9\alpha_{-2}^{L}\alpha_{-1}^{L}%
+9\alpha_{-2}^{T}\alpha_{-1}^{T}+75\alpha_{-2}^{P}\alpha_{-1}^{P}%
+50\alpha_{-3}^{P}\right]  \left\vert 0,k\right\rangle . \label{R4.1}%
\end{equation}
For the two type I spin ZNSs in Equation (\ref{42}), we define%
\begin{equation}
\theta_{\mu\nu}=\sum_{\alpha,\beta}e_{\mu}^{\alpha}e_{\nu}^{\beta}%
u_{\alpha\beta}.
\end{equation}
The transverse and traceless conditions on $\theta_{\mu\nu}$ then imply%
\begin{equation}
u_{PP}=u_{PL}=u_{PT}=0\text{ and }u_{PP}-u_{LL}-u_{TT}=0, \label{MM}%
\end{equation}
which gives two LZNSs:
\begin{align}
(\alpha_{-1}^{L}\alpha_{-2}^{L}+\alpha_{-1}^{P}\alpha_{-1}^{L}\alpha_{-1}%
^{L}-\alpha_{-1}^{T}\alpha_{-2}^{T}-\alpha_{-1}^{P}\alpha_{-1}^{T}\alpha
_{-1}^{T})|0,k\rangle &  ,\\
(\alpha_{-1}^{(L}\alpha_{-2}^{T)}+\alpha_{-1}^{P}\alpha_{-1}^{L}\alpha
_{-1}^{T})|0,k\rangle &  .
\end{align}

The vector ZNS in Equation (\ref{43}) gives two LZNSs:
\begin{equation}
\lbrack6\alpha_{-3}^{T}+18\alpha_{-1}^{(P}\alpha_{-2}^{T)}+9\alpha_{-1}%
^{P}\alpha_{-1}^{P}\alpha_{-1}^{T}+\alpha_{-1}^{L}\alpha_{-1}^{L}\alpha
_{-1}^{T}+\alpha_{-1}^{T}\alpha_{-1}^{T}\alpha_{-1}^{T}]|0,k\rangle,
\end{equation}%
\begin{equation}
\lbrack6\alpha_{-3}^{L}+18\alpha_{-1}^{(P}\alpha_{-2}^{L)}+9\alpha_{-1}%
^{P}\alpha_{-1}^{P}\alpha_{-1}^{L}+\alpha_{-1}^{L}\alpha_{-1}^{L}\alpha
_{-1}^{L}+\alpha_{-1}^{L}\alpha_{-1}^{T}\alpha_{-1}^{T}]|0,k\rangle.
\end{equation}
\ The vector ZNS in Equation (\ref{44}) gives two LZNS:
\begin{equation}
\lbrack3\alpha_{-3}^{T}-9\alpha_{-1}^{[P}\alpha_{-2}^{T]}-\alpha_{-1}%
^{L}\alpha_{-1}^{L}\alpha_{-1}^{T}-\alpha_{-1}^{T}\alpha_{-1}^{T}\alpha
_{-1}^{T}]|0,k\rangle,
\end{equation}%
\begin{equation}
\lbrack3\alpha_{-3}^{L}-9\alpha_{-1}^{[P}\alpha_{-2}^{L]}-\alpha_{-1}%
^{L}\alpha_{-1}^{L}\alpha_{-1}^{L}-\alpha_{-1}^{L}\alpha_{-1}^{T}\alpha
_{-1}^{T}]|0,k\rangle.
\end{equation}

In conclusion, there are seven LZNSs in total at the mass level $M^{2}=$ $4$.

It is important to note that there are nine LSSAs at mass level $M^{2}=$ $2$
with only three LZNSs, and $22$ LSSAs at mass level $M^{2}=$ $4$ with only seven
LZNSs. Thus, in contrast to the recurrence relations calculated in Equations (\ref{key}) and  (\ref{M}), these Ward identities are not enough to solve all the LSSAs
and express them in terms of one amplitude.

\subsubsection{The Lauricella Ward Identities}

In this subsection, we explicitly verify some examples of Ward identities
through processes (A),(B) and (C). Process (C) is implicitly used through
the kinematics. Ward identities cannot be independent of
the recurrence relations used in processes (A),(B) and (C) in the last section.

For $M^{2}=$ $2$, we define the following kinematics variables:
\begin{align}
\alpha &  =\frac{-t}{2}-1=Mk_{3}^{P}-N+1=\sqrt{2}k_{3}^{P}-1,\\
\gamma &  =\frac{s}{2}+2-N=-Mk_{1}^{P}=-\sqrt{2}k_{1}^{P},\\
d  &  =\left(  \frac{-k_{1}^{L}}{k_{3}^{L}}\right)  ^{\frac{1}{2}},1-\left(
\frac{-k_{1}^{P}}{k_{3}^{P}}\right)  =\frac{\alpha-\gamma+1}{\alpha+1},
\end{align}
then%
\begin{equation}
\frac{u}{2}+2-N=\alpha-\gamma+1-N=\alpha-\gamma-1.
\end{equation}

As examples, we calculate the Ward identities associated with the LZNSs in
\linebreak Equations (\ref{R2.1}) and (\ref{R2.2}). The calculation is based on processes (A)
and (B). By using Equation (\ref{st1}), the Ward identities we want to prove are\vspace{6pt}
\begin{align}
\left(  -k_{3}^{T}\right)  F_{D}^{(2)}\left(  \alpha;-1,-1;\alpha
-\gamma-1;1-\left(  \frac{-k_{1}^{T}}{k_{3}^{T}}\right)  ^{\frac{1}{2}%
},1+\left(  \frac{-k_{1}^{T}}{k_{3}^{T}}\right)  ^{\frac{1}{2}}\right)   &
\nonumber\\
+\sqrt{2}\left(  -k_{3}^{P}\right)  \left(  -k_{3}^{T}\right)  F_{D}%
^{(2)}\left(  \alpha;-1,-1;\alpha-\gamma-1;1-\left(  \frac{-k_{1}^{P}}%
{k_{3}^{P}}\right)  ,1-\left(  \frac{-k_{1}^{T}}{k_{3}^{T}}\right)  \right)
&  =0,\\
\left(  -k_{3}^{L}\right)  F_{D}^{(2)}\left(  \alpha;-1,-1;\alpha
-\gamma-1;1-\left(  \frac{-k_{1}^{L}}{k_{3}^{L}}\right)  ^{\frac{1}{2}%
},1+\left(  \frac{-k_{1}^{L}}{k_{3}^{L}}\right)  ^{\frac{1}{2}}\right)   &
\nonumber\\
+\sqrt{2}\left(  -k_{3}^{P}\right)  \left(  -k_{3}^{L}\right)  F_{D}%
^{(2)}\left(  \alpha;-1,-1;\alpha-\gamma-1;1-\left(  \frac{-k_{1}^{P}}%
{k_{3}^{P}}\right)  ,1-\left(  \frac{-k_{1}^{L}}{k_{3}^{L}}\right)  \right)
&  =0
\end{align}
or, using the kinematics variables just defined,
\begin{align}
F_{D}^{(2)}(a;-1,-1;\alpha-\gamma-1;1,1)-(\alpha+1)F_{D}^{(2)}\left(
\alpha;-1,-1;\alpha-\gamma-1;\frac{\alpha-\gamma+1}{\alpha+1},1\right)   &
=0,\label{W1}\\
F_{D}^{(2)}(\alpha;-1,-1;\alpha-\gamma-1;1-d,1+d)-(\alpha+1)F_{D}^{(2)}\left(
\alpha;-1,-1;\alpha-\gamma-1;\frac{\alpha-\gamma+1}{\alpha+1},1-d^{2}\right)
&  =0. \label{W2}%
\end{align}
Equations (\ref{W1}) and  (\ref{W2}) can be explicitly proved as
\begin{align}
&  F_{D}^{(2)}(\alpha;-1,-1;\alpha-\gamma-1;1,1)-(\alpha+1)F_{D}^{(2)}\left(
\alpha;-1,-1;\alpha-\gamma-1;\frac{\alpha-\gamma+1}{\alpha+1},1\right)
\nonumber\\
&  =F_{D}^{(1)}(\alpha;-2;\alpha-\gamma-1;1)-(\alpha+1)\left[
\begin{array}
[c]{c}%
\frac{\alpha-\gamma+1}{\alpha+1}F_{D}^{(1)}\left(  \alpha;-2;\alpha
-\gamma-1;1\right) \\
+\frac{\gamma}{\alpha+1}F_{D}^{(1)}\left(  \alpha;-1;\alpha-\gamma-1;1\right)
\end{array}
\right] \label{WW1}\\
&  =(\gamma-\alpha)F_{D}^{(1)}\left(  \alpha;-2;\alpha-\gamma-1;1\right)
-\gamma F_{D}^{(1)}\left(  \alpha;-1;\alpha-\gamma-1;1\right) \nonumber\\
&  =0, \label{W10}%
\end{align}
and
\begin{align}
&  F_{D}^{(2)}(\alpha;-1,-1;\alpha-\gamma-1;1-d,1+d)-(\alpha+1)F_{D}%
^{(2)}\left(  \alpha;-1,-1;\alpha-\gamma-1;\frac{\alpha-\gamma+1}{\alpha
+1},1-d^{2}\right) \nonumber\\
&  =\frac{1-d}{1+d}F_{D}^{(1)}(\alpha;-2;\alpha-\gamma-1;1+d)-\frac{2d}%
{1+d}F_{D}^{(1)}(\alpha;-1;\alpha-\gamma-1,1+d)\nonumber\\
&  -(\alpha+1)\left[
\begin{array}
[c]{c}%
\frac{\alpha-\gamma+1}{(\alpha+1)(1-d^{2})}F_{D}^{(1)}\left(  \alpha
;-2;\alpha-\gamma-1;1-d^{2}\right) \\
+\left(  \frac{\alpha-\gamma+1}{(\alpha+1)(1-d^{2})}-(1-d)\right)  F_{D}%
^{(1)}(\alpha;-1;\alpha-\gamma-1;1-d^{2})
\end{array}
\right] \label{WW2}\\
&  =\frac{1-d}{1+d}\left(  1-\frac{2\alpha d}{\gamma-1}+\frac{\alpha
(\alpha+1)^{2}}{(\gamma-1)(\gamma-2)}\right)  F_{D}^{(1)}(\alpha
;-2;\alpha-\gamma-1;1)\nonumber\\
&  -\frac{2d}{1+d}\left(  1-\frac{\alpha d}{\gamma}\right)  F_{D}^{(1)}%
(\alpha;-1;\alpha-\gamma-1;1)\nonumber\\
&  -(\alpha+1)\left[
\begin{array}
[c]{c}%
\frac{\alpha-\gamma+1}{(\alpha+1)(1-d^{2})}\left(  1+\frac{2\alpha d^{2}%
}{\gamma-1}+\frac{\alpha(\alpha+1)d^{4}}{(\gamma-1)(\gamma-2)}\right)
F_{D}^{(1)}(\alpha;-2;\alpha-\gamma-1;1)\\
+\left(  \frac{\alpha-\gamma+1}{(\alpha+1)(1-d^{2})}-(1-d)\right)  \left(
1+\frac{\alpha d^{2}}{\gamma}\right)  F_{D}^{(1)}(\alpha;-1;\alpha-\gamma-1;1)
\end{array}
\right] \label{WWW2}\\
&  =0, \label{W20}%
\end{align}
where we used Equation (\ref{key}) in process (A) to get Equations (\ref{WW1}) and (\ref{WW2}) and Equation (\ref{F}) in process (B) to get Equation (\ref{WWW2}).
The last last lines of the above equations were obtained by using Equation (\ref{F4}).

\subsection{Summary}

In this section, we have shown that there is an infinite number of recurrence
relations valid for all energies among the LSSA of three tachyons and
one arbitrary string state. Moreover, this infinite number of recurrence
relations can be used to solve all the LSSAs and express them in terms of one
single four tachyon amplitude. In addition, we find that the $K+2$ recurrence
relations among the LSSA can be used to reproduce the Cartan subalgebra and
simple root system of the $SL(K+3,%
\mathbb{C}
)$ group with rank $K+2$. Thus, the recurrence relations are equivalent to the
representation of $SL(K+3,%
\mathbb{C}
)$ group of the LSSA. As a result, the $SL(K+3,%
\mathbb{C}
)$ group can be used to solve all LSSAs and express them in terms of one
amplitude \cite{1707.01281}.

We have also shown that, for the first few mass levels, the solvability of LSSAs
through the calculations of recurrence relations implies the validity of Ward
identities derived from the decoupling of LZNS. However, the Lauricella Ward
identities are not good enough to solve all the LSSAs and express them
in terms of one amplitude.%


\section{Relations among LSSAs in Various Scattering Limits}

In this section, we show that there exist relations or symmetries among
SSAs of different string states at various scattering limits. In the first
subsection, we show that the linear relations
\cite{GM,GM1,Gross,Gross1,GrossManes} conjectured by Gross among the hard SSAs
(HSSAs) at each fixed mass level in the hard scattering limit can be rederived
from the LSSA. These relations reduce the number of independent HSSAs from
$\infty$ down to $1$.

In the second subsection, we show that the Regge SSA (RSSA) in the Regge
scattering limit can be rederived from the LSSA. All the RSSAs can be expressed
in terms of the Appell functions with associated $SL(5,%
\mathbb{C}
)$ symmetry \cite{KLY,LY,AppellLY}. Moreover, the recurrence relations of the
Appell functions can be used to reduce the number of independent RSSAs from
$\infty$ down to $1$.

Finally, in the nonrelativistic scattering limit, we show that the
nonrelativistic SSAs (NSSAs) and various extended recurrence relations among
them an be rederived from the LSSA. In addition, we also derive the
nonrelativistic level $M_{2}$-dependent string BCJ relations, which
are the stringy generalization of the massless field theory BCJ relation
\cite{BCJ1} to the higher spin stringy particles. These NSSAs can be expressed
in terms of the Gauss hypergeometry functions with associated $SL(4,%
\mathbb{C}
)$ symmetry \cite{KLY,LY,AppellLY}.

\subsection{Hard Scattering Limit---Proving the Gross Conjecture from LSSAs}

In this subsection, we show that the linear relations conjectured by
Gross \cite{GM,GM1,Gross,Gross1,GrossManes} in the hard scattering limit can
be rederived from the LSSA. First, we briefly review the results discussed in
\cite{review,over} for the linear relations among HSSAs. It was first observed
that for each fixed mass level $N$ with $M^{2}=2(N-1)$, the following states
are of a leading order in energy at the hard scattering limit
\cite{CHLTY2,CHLTY1}
\begin{equation}
\left\vert N,2m,q\right\rangle \equiv(\alpha_{-1}^{T})^{N-2m-2q}(\alpha
_{-1}^{L})^{2m}(\alpha_{-2}^{L})^{q}|0,k\rangle. \label{Nmq}%
\end{equation}
Note that in Equation (\ref{Nmq}), only even powers $2m$ in $\alpha_{-1}^{L}$
\cite{ChanLee1,ChanLee,ChanLee2} survive, and the naive energy order of the
amplitudes will drop by an even number of energy powers in general. The HSSAs
with vertices corresponding to states with an odd power in $(\alpha_{-1}%
^{L})^{2m+1}$\ turn out to be of a subleading order in energy and can be
ignored. By using the stringy Ward identities or the decoupling of two types of
zero norm states (ZNSs) in the hard scattering limit, the linear relations
among HSSAs of different string states at each fixed mass level $N$ were
calculated to be \cite{CHLTY2,CHLTY1}
\begin{equation}
\frac{A_{st}^{(N,2m,q)}}{A_{st}^{(N,0,0)}}=\left(  -\frac{1}{M}\right)
^{2m+q}\left(  \frac{1}{2}\right)  ^{m+q}(2m-1)!!. \label{04}%
\end{equation}
Exactly the same result can be obtained by using two other techniques: the
Virasoro constraint calculation and the corrected saddle-point calculation
\cite{CHLTY2,CHLTY1}. The calculation of of Equation (\ref{04}) was first done for
one high-energy vertex in Equation (\ref{Nmq}) and could then be easily generalized to
four high-energy vertices. In the decoupling of ZNS calculations at the mass
level $M^{2}=4$, for example, there are four leading order HSSAs
\cite{ChanLee1,ChanLee2}
\begin{equation}
A_{TTT}:A_{LLT}:A_{(LT)}:A_{[LT]}=8:1:-1:-1 \label{03}%
\end{equation}
which are proportional to each other. However. the saddle point calculation of
\cite{GrossManes} gave $A_{TTT}\propto A_{[LT]},$ and $A_{LLT}=0$, which are
inconsistent with the decoupling of ZNS or unitarity of the theory. Indeed, a
sample calculation was done \cite{ChanLee1,ChanLee2} to explicitly verify the
ratios in Equation (\ref{03}).

One interesting application of Equation (\ref{04}) was the derivation of the ratio
between $A_{st}^{(N,2m,q)}$ and $A_{tu}^{(N,2m,q)}$ in the hard scattering
limit \cite{Closed}%
\begin{equation}
A_{st}^{(N,2m,q)}\simeq(-)^{N}\frac{\sin(\pi k_{2}\cdot k_{4})}{\sin(\pi
k_{1}\cdot k_{2})}A_{tu}^{(N,2m,q)} \label{HBCJ}%
\end{equation}
where $A_{tu}^{(N,2m,q)}$ is the corresponding $(t,u)$ channel HSSA.

Equation (\ref{HBCJ}) was shown to be valid for scatterings of four arbitrary string
states in the hard scattering limit and was obtained in 2006. This result was
obtained earlier than the discovery of four-point field theory BCJ relations
in \cite{BCJ1} and ``string BCJ relations'' in Equation (\ref{BCJ})
\cite{LLY,stringBCJ,stringBCJ2}. In contrast to the the calculation of string
BCJ relations in \cite{stringBCJ, stringBCJ2}, which was motivated by the field
theory BCJ relations in \cite{BCJ1}, the result of Equation (\ref{HBCJ}) was
inspired by the calculation of hard closed SSAs \cite{Closed} by using
the KLT 
relation \cite{KLT}. More detailed discussion can be found in
\cite{Closed,over}.

Thus, we are ready to rederive Equations (\ref{Nmq}) and (\ref{04}) from the LSSA in \linebreak
Equation (\ref{st1}). The relevant kinematics are%
\begin{align}
k_{1}^{T}  &  =0\text{, \ \ }k_{3}^{T}\simeq-E\sin\phi,\\
k_{1}^{L}  &  \simeq-\frac{2p^{2}}{M_{2}}\simeq-\frac{2E^{2}}{M_{2}},\\
k_{3}^{L}  &  \simeq\frac{2E^{2}}{M_{2}}\sin^{2}\frac{\phi}{2}.
\end{align}
where $E$ and $\phi$ are the CM frame energy and scattering angle, respectively.
One can calculate%
\begin{equation}
\tilde{z}_{kk^{\prime}}^{T}=1,\ \tilde{z}_{kk^{\prime}}^{L}=1-\left(
-\frac{s}{t}\right)  ^{1/k}e^{\frac{i2\pi k^{\prime}}{k}}\sim O\left(
1\right)  .
\end{equation}
The LSSA in Equation (\ref{st1}) reduces to%
\begin{align}
&  A_{st}^{(r_{n}^{T},r_{l}^{L})}=B\left(  -\frac{t}{2}-1,-\frac{s}%
{2}-1\right) \nonumber\\
&  \cdot\prod_{n=1}\left[  (n-1)!E\sin\phi\right]  ^{r_{n}^{T}}\prod
_{l=1}\left[  -(l-1)!\frac{2E^{2}}{M_{2}}\sin^{2}\frac{\phi}{2}\right]
^{r_{l}^{L}}\nonumber\\
&  \cdot F_{D}^{(K)}\left(  -\frac{t}{2}-1;R_{n}^{T},R_{l}^{L};\frac{u}%
{2}+2-N;\left(  1\right)  _{n},\tilde{Z}_{l}^{L}\right)  .
\end{align}

As mentioned above, in the hard scattering limit, there was a
difference between the naive energy order and the real energy order
corresponding to the $\left(  \alpha_{-1}^{L}\right)  ^{r_{1}^{L}}$ operator
in Equation (\ref{state}). Thus, it is important to pay attention to the corresponding summation and
write
\vspace{6pt}
\begin{align}
&  A_{st}^{(r_{n}^{T},r_{l}^{L})}=B\left(  -\frac{t}{2}-1,-\frac{s}%
{2}-1\right) \nonumber\\
&  \cdot\prod_{n=1}\left[  (n-1)!E\sin\phi\right]  ^{r_{n}^{T}}\prod
_{l=1}\left[  -(l-1)!\frac{2E^{2}}{M_{2}}\sin^{2}\frac{\phi}{2}\right]
^{r_{l}^{L}}\nonumber\\
&  \cdot\sum_{k_{r}}\frac{\left(  -\frac{t}{2}-1\right)  _{k_{r}}}{\left(
\frac{u}{2}+2-N\right)  _{k_{r}}}\frac{\left(  -r_{1}^{L}\right)  _{k_{r}}%
}{k_{r}!}\left(  1+\frac{s}{t}\right)  ^{k_{r}}\cdot\left(  \cdots\right)
\end{align}
where  $\left(  a\right)  _{n+m}=\left(  a\right)  _{n}\left(
a+n\right)  _{m}$ and $\left(  \cdots\right)  $ are terms which are not
relevant to the following discussion. We then propose the following formula:
\begin{align}
&  \sum_{k_{r}=0}^{r_{1}^{L}}\frac{\left(  -\frac{t}{2}-1\right)  _{k_{r}}%
}{\left(  \frac{u}{2}+2-N\right)  _{k_{r}}}\frac{\left(  -r_{1}^{L}\right)
_{k_{r}}}{k_{r}!}\left(  1+\frac{s}{t}\right)  ^{k_{r}}\nonumber\\
=  &  0\cdot\left(  \frac{tu}{s}\right)  ^{0}\!+0\cdot\left(  \frac{tu}%
{s}\right)  ^{-1}\!+\dots+0\cdot\left(  \frac{tu}{s}\right)  ^{-\left[
\frac{r_{1}^{L}+1}{2}\right]  -1}\nonumber\\
&  +C_{r_{1}^{L}}\left(  \frac{tu}{s}\right)  ^{-\left[  \frac{r_{1}^{L}+1}%
{2}\right]  }+\mathit{O}\left\{  \left(  \frac{tu}{s}\right)  ^{-\left[
\frac{r_{1}^{L}+1}{2}\right]  +1}\right\}  . \label{pro}%
\end{align}
where $[$ $]$ stands for the Gauss symbol, $C_{r_{1}^{L}}$ is independent of
energy $E$ and depends on $r_{1}^{L}$ and possibly the scattering angle $\phi
$. When $r_{1}^{L}=2m$\ is an even number, we further propose that
$C_{r_{1}^{L}}=\frac{\left(  2m\right)  !}{m!}$ and is $\phi$ independent. We
have verified Equation (\ref{pro}) for $r_{1}^{L}=0,1,2,\cdots,10$.

Notice that Equation (\ref{pro}) reduces to the Stirling number identity by taking
the Regge limit ($s\rightarrow\infty$ with $t$ fixed) and setting $r_{1}%
^{L}=2m$,
\begin{gather}
\sum_{k_{r}=0}^{2m}\frac{\left(  -\frac{t}{2}-1\right)  _{k_{r}}}{\left(
-\frac{s}{2}\right)  _{k_{r}}}\frac{\left(  -2m\right)  _{k_{r}}}{k_{r}
!}\left(  \frac{s}{t}\right)  ^{k_{r}}\simeq\sum_{k_{r}=0}^{2m}\left(
-2m\right)  _{k_{r}}\left(  -\frac{t}{2}-1\right)  _{k_{r}}\frac{\left(
-2/t\right)  ^{k_{r}}}{k_{r}!}\nonumber\\
=0\cdot\left(  -t\right)  ^{0}\!+0\cdot\left(  -t\right)  ^{-1}\!+\dots
+0\cdot\left(  -t\right)  ^{-m+1}+\frac{(2m)!}{m!}\left(  -t\right)
^{-m}+\mathit{O}\left\{  \left(  \frac{1}{t}\right)  ^{m+1}\right\}  ,
\label{stirling}%
\end{gather}
which was proposed in \cite{KLY} and proved in \cite{LYAM}.

It was demonstrated in \cite{KLY} that the ratios in the hard scattering limit
in\linebreak Equation (\ref{04}) can be reproduced from a class of Regge string scattering
amplitudes presented in Equation (\ref{app}). The key of the proof of this
relationship between HSSA and RSSA was the new Stirling number identity
proposed in Equation  (\ref{stirling}) and mathematical proved in \cite{LYAM}. On the
other hand, the mathematical proof of Equation (\ref{pro}), which is a
generalization of the identity in Equation (\ref{stirling}), is an open question and
may be an interesting one to study.

The zero terms in Equation (\ref{pro}) correspond to the naive leading energy orders in
the HSSA calculation. In the hard scattering limit, the true leading order SSA
can then be~identified:
\vspace{6pt}
\begin{align}
&  A_{st}^{(r_{n}^{T},r_{l}^{L})}\simeq B\left(  -\frac{t}{2}-1,-\frac{s}%
{2}-1\right) \nonumber\\
&  \cdot\prod_{n=1}\left[  (n-1)!E\sin\phi\right]  ^{r_{n}^{T}}\prod
_{l=1}\left[  -(l-1)!\frac{2E^{2}}{M_{2}}\sin^{2}\frac{\phi}{2}\right]
^{r_{l}^{L}}\nonumber\\
&  \cdot C_{r_{1}^{L}}\left(  E\sin\phi\right)  ^{-2\left[  \frac{r_{1}^{L}%
+1}{2}\right]  }\cdot\left(  \cdots\right) \nonumber\\
&  \sim E^{N-\sum_{n\geq2}nr_{n}^{T}-\left(  2\left[  \frac{r_{1}^{L}+1}%
{2}\right]  -r_{1}^{L}\right)  -\sum_{l\geq3}lr_{l}^{L}},
\end{align}
which means that SSA reaches its highest energy when $r_{n\geq2}^{T}%
=r_{l\geq3}^{L}=0$ and $r_{1}^{L}=2m$---an even number. This result is
consistent with the previous result presented in \linebreak Equation (\ref{Nmq})
\cite{ChanLee1,ChanLee,ChanLee2,CHL,CHLTY2, CHLTY1,susy}.

Finally, the leading order SSA in the hard scattering limit, i.e., $r_{1}%
^{T}=N-2m-2q$, $r_{1}^{L}=2m$ and $r_{2}^{L}=q$, can be calculated to be%
\begin{align}
&  A_{st}^{(N-2m-2q,2m,q)}\nonumber\\
&  \simeq B\left(  -\frac{t}{2}-1,-\frac{s}{2}-1\right)  \left(  E\sin
\phi\right)  ^{N}\frac{\left(  2m\right)  !}{m!}\left(  -\frac{1}{2M_{2}%
}\right)  ^{2m+q}\nonumber\\
&  =(2m-1)!!\left(  -\frac{1}{M_{2}}\right)  ^{2m+q}\left(  \frac{1}%
{2}\right)  ^{m+q}A_{st}^{(N,0,0)}%
\end{align}
which reproduces the ratios in Equation (\ref{04}), and is consistent with the
previous\linebreak result \cite{ChanLee1,ChanLee,ChanLee2,CHL,CHLTY2, CHLTY1,susy}.

\subsection{Regge Scattering Limit}
  There is another important high-energy limit of SSA: the RSSA in the
Regge scattering limit. The relevant kinematics in the Regge limit are%
\begin{align}
k_{1}^{T}  &  =0\text{, \ \ }k_{3}^{T}\simeq-\sqrt{-t},\\
k_{1}^{P}  &  \simeq-\frac{s}{2M_{2}}\text{,\ }k_{3}^{P}\simeq-\frac{\tilde
{t}}{2M_{2}}=-\frac{t-M_{2}^{2}-M_{3}^{2}}{2M_{2}},\\
k_{1}^{L}  &  \simeq-\frac{s}{2M_{2}}\text{, }k_{3}^{L}\simeq-\frac{\tilde
{t}^{\prime}}{2M_{2}}=-\frac{t+M_{2}^{2}-M_{3}^{2}}{2M_{2}}.
\end{align}
One can easily calculate%
\begin{equation}
\tilde{z}_{kk^{\prime}}^{T}=1,\ \tilde{z}_{kk^{\prime}}^{P}=1-\left(
-\frac{s}{\tilde{t}}\right)  ^{1/k}e^{\frac{i2\pi k^{\prime}}{k}}\sim s^{1/k}%
\end{equation}
and%
\begin{equation}
\tilde{z}_{kk^{\prime}}^{L}=1-\left(  -\frac{s}{\tilde{t}^{\prime}}\right)
^{1/k}e^{\frac{i2\pi k^{\prime}}{k}}\sim s^{1/k}.
\end{equation}
In the Regge limit, the SSA in Equation (\ref{st2}) reduces to%
\begin{align}
&  A_{st}^{(r_{n}^{T},r_{m}^{P},r_{l}^{L})}\nonumber\\
\simeq &  B\left(  -\frac{t}{2}-1,-\frac{s}{2}-1\right)  \prod_{n=1}\left[
(n-1)!\sqrt{-t}\right]  ^{r_{n}^{T}}\nonumber\\
\cdot &  \prod_{m=1}\left[  (m-1)!\frac{\tilde{t}}{2M_{2}}\right]  ^{r_{m}%
^{P}}\prod_{l=1}\left[  (l-1)!\frac{\tilde{t}^{\prime}}{2M_{2}}\right]
^{r_{l}^{L}}\nonumber\\
\cdot &  F_{1}\left(  -\frac{t}{2}-1;-q_{1},-r_{1};-\frac{s}{2};\frac
{s}{\tilde{t}},\frac{s}{\tilde{t}^{\prime}}\right)  . \label{app}%
\end{align}
where $F_{1}$ is the Appell function. Equation (\ref{app}) agrees with the result
obtained in \cite{AppellLY} previously.

The recurrence relations of the Appell functions can be used to reduce the
number of independent RSSAs from $\infty$ down to $1$. One can also calculate
the string BCJ relation in the Regge scattering limit and study the extended
recurrence relation in the Regge limit \cite{LLY}.

\subsection{Nonrelativistic Scattering Limit and Extended Recurrence
Relations}

In this section, we discuss nonrelativistic string scattering amplitudes
(NSSAs) and the extended recurrence relations among them. In addition, we will
also derive the nonrelativistic level $M_{2}$-dependent string BCJ
relations which are the stringy generalization of the massless field
theory BCJ relation \cite{BCJ1} to the higher spin stringy particles.

We employ the nonrelativistic string scattering limit or $|\vec{k_{2}%
}|<<M_{2}$ limit to calculate the mass level and spin dependent low-energy
SSA. In contrast to the zero slope $\alpha^{\prime}$ limit used in the
literature to calculate the massless Yang--Mills couplings
\cite{ymzero1,ymzero2} for superstrings and the three point $\varphi^{3}$
scalar field coupling \cite{Bzero1,Bzero2,Bzero3} for bosonic strings, we
found it appropriate to take the nonrelativistic limit to calculate low-energy SSAs for string states with both higher spins and finite mass gaps.

\subsubsection{Nonrelavistic LSSA}

  In this subsection, we first calculate the NSSA from the LSSA. In the
nonrelativistic limit $|\vec{k_{1}}|\ll M_{2}$, we have%
\begin{align}
k_{1}^{T}  &  =0,k_{3}^{T}=-\left[  \frac{\epsilon}{2}+\frac{(M_{1}+M_{2}%
)^{2}}{4M_{1}M_{2}\epsilon}|\vec{k_{1}}|^{2}\right]  \sin\phi,\\
k_{1}^{L}  &  =-\frac{M_{1}+M_{2}}{M_{2}}|\vec{k_{1}}|+O\left(  |\vec{k_{1}%
}|^{2}\right)  ,\\
k_{3}^{L}  &  =-\frac{\epsilon}{2}\cos\phi+\frac{M_{1}+M_{2}}{2M_{2}}%
|\vec{k_{1}}|+O\left(  |\vec{k_{1}}|^{2}\right)  ,\\
k_{1}^{P}  &  =-M_{1}+O\left(  |\vec{k_{1}}|^{2}\right)  ,\\
k_{3}^{P}  &  =\frac{M_{1}+M_{2}}{2}-\frac{\epsilon}{2M_{2}}\cos\phi
|\vec{k_{1}}|+O\left(  |\vec{k_{1}}|^{2}\right)
\end{align}
where $\epsilon=\sqrt{(M_{1}+M_{2})^{2}-4M_{3}^{2}}$ and $M_{1}=M_{3}%
=M_{4}=M_{tachyon}$. One can easily calculate%
\begin{equation}
z_{k}^{T}=z_{k}^{L}=0,z_{k}^{P}\simeq\left\vert \left(  \frac{2M_{1}}%
{M_{1}+M_{2}}\right)  ^{\frac{1}{k}}\right\vert .
\end{equation}
The SSA in Equation (\ref{st2}) reduces to%
\begin{align}
&  A_{st}^{(r_{n}^{T},r_{m}^{P},r_{l}^{L})}\nonumber\\
&  \simeq\prod_{n=1}\left[  (n-1)!\frac{\epsilon}{2}\sin\phi\right]
^{r_{n}^{T}}\prod_{m=1}\left[  -(m-1)!\frac{M_{1}+M_{2}}{2}\right]
^{r_{m}^{P}}\nonumber\\
&  \cdot\prod_{l=1}\left[  (l-1)!\frac{\epsilon}{2}\cos\phi\right]
^{r_{l}^{L}}B\left(  \frac{M_{1}M_{2}}{2},1-M_{1}M_{2}\right) \nonumber\\
&  \cdot F_{D}^{(K)}\left(  \frac{M_{1}M_{2}}{2};R_{m}^{P};M_{1}M_{2};\left(
\frac{2M_{1}}{M_{1}+M_{2}}\right)  _{m}\right)
\end{align}
where%
\begin{equation}
K=\underset{\{\text{for all }r_{m}^{P}\neq0\}}{\sum m}.
\end{equation}

\subsubsection{Nonrelativistic String BCJ Relations}

Note that for string states with $r_{k}^{P}=0$ in Equation (\ref{st2}) for all
$k\geq2$, one has $K=1$, and the Lauricella functions in the low-energy
nonrelativistic SSA reduce to the Gauss hypergeometric functions $F_{D}%
^{(1)}=$ $_{2}F_{1}$ with the associated $SL(4,C)$ symmetry. In particular,
for the case of the leading trajectory string state in the second vertex with
mass level $N=N_{1}+N_{2}+N_{3}$ where $r_{1}^{T}=N_{1}$, $r_{1}^{P}=N_{3}$,
$r_{1}^{L}=N_{2}$, and $r_{k}^{X}=0$ for all $k\geq2$, the SSA reduces to%
\begin{align}
&  A_{st}^{(N_{1},N_{2},N_{3})}=\left(  \frac{\epsilon}{2}\sin\phi\right)
^{N_{1}}\left(  \frac{\epsilon}{2}\cos\phi\right)  ^{N_{2}}\nonumber\\
\cdot &  \left(  -\frac{M_{1}+M_{2}}{2}\right)  ^{N_{3}}B\left(  \frac
{M_{1}M_{2}}{2},1-M_{1}M_{2}\right) \nonumber\\
\cdot &  _{2}F_{1}\left(  \frac{M_{1}M_{2}}{2};-N_{3};M_{1}M_{2};\frac{2M_{1}%
}{M_{1}+M_{2}}\right)  , \label{low}%
\end{align}
which agrees with the result obtained in \cite{LLY} previously. Similarly, one
can calculate the corresponding nonrelativistic $t-u$ channel amplitude as%
\begin{align}
A_{tu}^{(N_{1},N_{2},N_{3})}=  &  \left(  -1\right)  ^{N}\left(
\frac{\epsilon}{2}\sin\phi\right)  ^{N_{1}}\left(  \frac{\epsilon}{2}\cos
\phi\right)  ^{N_{2}}\nonumber\\
&  \cdot\left(  -\frac{M_{1}+M_{2}}{2}\right)  ^{N_{3}}B\left(  \frac
{M_{1}M_{2}}{2},\frac{M_{1}M_{2}}{2}\right) \nonumber\\
&  \cdot\text{ }_{2}F_{1}\left(  \frac{M_{1}M_{2}}{2};-N_{3};M_{1}M_{2}%
;\frac{2M_{1}}{M_{1}+M_{2}}\right)  .
\end{align}
Finally, the ratio of $s-t$ and $t-u$ channel amplitudes is \cite{LLY}%

\begin{align}
\frac{A_{st}^{(N_{1},N_{2},N_{3})}}{A_{tu}^{(N_{1},N_{2},N_{3})}}  &  =\left(
-1\right)  ^{N}\frac{B\left(  -M_{1}M_{2}+1,\frac{M_{1}M_{2}}{2}\right)
}{B\left(  \frac{M_{1}M_{2}}{2},\frac{M_{1}M_{2}}{2}\right)  }\nonumber\\
&  =(-1)^{N}\frac{\Gamma\left(  M_{1}M_{2}\right)  \Gamma\left(  -M_{1}%
M_{2}+1\right)  }{\Gamma\left(  \frac{M_{1}M_{2}}{2}\right)  \Gamma\left(
-\frac{M_{1}M_{2}}{2}+1\right)  }\simeq\frac{\sin\pi\left(  k_{2}\cdot
k_{4}\right)  }{\sin\pi\left(  k_{1}\cdot k_{2}\right)  } \label{NBCJ}%
\end{align}
where, in the nonrelativistic limit, we have%
\begin{subequations}
\begin{align}
k_{1}\cdot k_{2}  &  \simeq-M_{1}M_{2},\\
k_{2}\cdot k_{4}  &  \simeq\frac{\left(  M_{1}+M_{2}\right)  M_{2}}{2}.
\end{align}

We thus obtain consistent nonrelativistic level $M_{2}$-dependent string BCJ relations. Similar relations for $t-u$ and
$s-u$ channel amplitudes can be calculated. We stress that the above relation
is the stringy generalization of the massless field theory BCJ relation
\cite{BCJ1} to the higher spin stringy particles. Moreover, as shown in
the next subsection, there are much more relations among the NSSAs.

\subsubsection{Extended Recurrence Relations in the Nonrelativistic Scattering
Limit}

\paragraph{Leading Trajectory String States}

In this subsection, we derive two examples of extended recurrence relations
among NSSAs. We first note that there is a recurrence relation of the Gauss
hypergeometry~function,
\end{subequations}
\begin{equation}
_{2}F_{1}(a;b;c;z)=\frac{c-2b+2+(b-a-1)z}{(b-1)(z-1)}\text{ }_{2}%
F_{1}(a;b-1;c;z)+\frac{b-c-1}{(b-1)(z-1)}\text{ }_{2}F_{1}(a;b-2;c;z),
\label{rec}%
\end{equation}
which can be used to derive the recurrence relation,
\begin{align}
\left(  -\frac{M_{1}+M_{2}}{2}\right)  A_{st}^{(p,r,q)}=  &  \frac
{M_{2}\left(  M_{1}M_{2}+2q+2\right)  }{\left(  q+1\right)  \left(
M_{2}-M_{1}\right)  }\left(  \frac{\epsilon}{2}\sin\phi\right)  ^{p-p^{\prime
}}\left(  \frac{\epsilon}{2}\cos\phi\right)  ^{p^{\prime}-p+1}A_{st}^{\left(
p^{\prime},p+r-p^{\prime}-1,q+1\right)  }\nonumber\\
+  &  \frac{2\left(  M_{1}M_{2}+q+1\right)  }{\left(  q+1\right)  \left(
M_{2}-M_{1}\right)  }\left(  \frac{\epsilon}{2}\sin\phi\right)  ^{p-p^{\prime
\prime}}\left(  \frac{\epsilon}{2}\cos\phi\right)  ^{p^{\prime\prime}%
-p+2}A_{st}^{\left(  p^{\prime\prime},p+r-p^{\prime\prime}-2,q+2\right)  }
\label{main}%
\end{align}
where $p^{\prime}$ and $p^{\prime\prime}$ are the polarization parameters of
the second and third amplitudes on the right-hand side of Equation (\ref{main}). For example,
for a fixed mass level $N=4$, one can derive many recurrence relations for
either $s-t$ channel or $t-u$ channel amplitudes with $q=0,1,2$. For example, for
$q=2,$ $(p,r)=(2,0),(1,1),(0,2)$, we have $p^{\prime}=0,1$ and $p^{\prime
\prime}=0.$ We can thus derive---for example, for $(p,r)=(2,0)$ and $p^{\prime
}=1$---the recurrence relation among amplitudes $A_{st}^{(2,0,2)}%
A_{st}^{(1,0,3)}A_{st}^{(0,0,4)}$ as follows:
\begin{equation}
\left(  -\frac{M_{1}+M_{2}}{2}\right)  A_{st}^{(2,0,2)}=\frac{M_{2}\left(
M_{1}M_{2}+6\right)  }{3\left(  M_{2}-M_{1}\right)  }\left(  \frac{\epsilon
}{2}\sin\phi\right)  A_{st}^{(1,0,3)}+\frac{2\left(  M_{1}M_{2}+4\right)
}{3\left(  M_{2}-M_{1}\right)  }\left(  \frac{\epsilon}{2}\sin\phi\right)
^{2}A_{st}^{(0,0,4)}.
\end{equation}
Exactly the same relation can be obtained for $t-u$ channel amplitudes since
the $_{2}F_{1}(a;b;c;z)$ dependence in the $s-t$ and $t-u$ channel amplitudes
calculated above are the same. Moreover, we can, for example, replace
the $A_{st}^{(2,0,2)}$ amplitude above by the corresponding $t-u$ channel
amplitude $A_{tu}^{(2,0,2)}$ through Equation (\ref{NBCJ}) and obtain%
\begin{align}
\frac{\left(  -1\right)  ^{N}}{2\cos\frac{\pi M_{1}M_{2}}{2}}\left(
-\frac{M_{1}+M_{2}}{2}\right)  A_{tu}^{(2,0,2)}  &  =\frac{M_{2}\left(
M_{1}M_{2}+6\right)  }{3\left(  M_{2}-M_{1}\right)  }\left(  \frac{\epsilon
}{2}\sin\phi\right)  A_{st}^{(1,0,3)}\nonumber\\
&  +\frac{2\left(  M_{1}M_{2}+4\right)  }{3\left(  M_{2}-M_{1}\right)
}\left(  \frac{\epsilon}{2}\sin\phi\right)  ^{2}A_{st}^{(0,0,4)}, \label{BCJJ}%
\end{align}
which relates higher spin nonrelativistic string amplitudes in both $s-t$ and
$t-u$ channels. Equation (\ref{BCJJ}) is one example of the extended
recurrence relations in the nonrelativistic string scattering~limit.

\paragraph{General String States}

Equation (\ref{BCJJ}) relates the NSSAs of different polarizations of a fixed leading
trajectory string state. In the next sample calculation, we calculate one
example of an extended recurrence relation that relates the NSS amplitudes of
different higher spin particles for each fixed mass level\ $M_{2}$. In
particular, the $s-t$ channel of the NSS amplitudes of three tachyons and one
higher spin massive string state at mass level $N=3p_{1}+q_{1}+3$
corresponding to the following three higher spin string states,
\begin{align}
&  A_{1}\symbol{126}\left(  i\partial^{3}X^{T}\right)  ^{p_{1}}\left(
i\partial X^{P}\right)  ^{1}\left(  i\partial X^{L}\right)  ^{q_{1}+2},\\
&  A_{2}\symbol{126}\left(  i\partial^{2}X^{T}\right)  ^{p_{1}}\left(
i\partial X^{P}\right)  ^{2}\left(  i\partial X^{L}\right)  ^{p_{1}+q_{1}%
+1},\\
&  A_{3}\symbol{126}\left(  i\partial X^{T}\right)  ^{p_{1}}\left(  i\partial
X^{P}\right)  ^{3}\left(  i\partial X^{L}\right)  ^{2p_{1}+q_{1}},
\end{align}
can be calculated to be%
\vspace{6pt}
\begin{align}
A_{1}  &  =\left[  2!\frac{\epsilon}{2}\sin\phi\right]  ^{p_{1}}\left[
-\left(  1-1\right)  !\frac{M_{1}+M_{2}}{2}\right]  ^{1}\left[  0!\frac
{\epsilon}{2}\cos\phi\right]  ^{q_{1}+2}\nonumber\\
&  \times B\left(  \frac{M_{1}M_{2}}{2},1-M_{1}M_{2}\right)  \text{ }_{2}%
F_{1}\left(  \frac{M_{1}M_{2}}{2},-1,M_{1}M_{2},\frac{-2M_{1}}{M_{1}+M_{2}%
}\right)  ,\\
A_{2}  &  =\left[  1!\frac{\epsilon}{2}\sin\phi\right]  ^{p_{1}}\left[
-\left(  2-1\right)  !\frac{M_{1}+M_{2}}{2}\right]  ^{2}\left[  0!\frac
{\epsilon}{2}\cos\phi\right]  ^{p_{1}+q_{1}+1}\nonumber\\
&  \times B\left(  \frac{M_{1}M_{2}}{2},1-M_{1}M_{2}\right)  \text{ }_{2}%
F_{1}\left(  \frac{M_{1}M_{2}}{2},-2,M_{1}M_{2},\frac{-2M_{1}}{M_{1}+M_{2}%
}\right)  ,\\
A_{3}  &  =\left[  0!\frac{\epsilon}{2}\sin\phi\right]  ^{p_{1}}\left[
-\left(  3-1\right)  !\frac{M_{1}+M_{2}}{2}\right]  ^{3}\left[  0!\frac
{\epsilon}{2}\cos\phi\right]  ^{2p_{1}+q_{1}}\nonumber\\
&  \times B\left(  \frac{M_{1}M_{2}}{2},1-M_{1}M_{2}\right)  \text{ }_{2}%
F_{1}\left(  \frac{M_{1}M_{2}}{2},-3,M_{1}M_{2},\frac{-2M_{1}}{M_{1}+M_{2}%
}\right)  .
\end{align}
To apply the recurrence relation in Equation (\ref{rec}) for Gauss hypergeometry
functions, we~choose%
\begin{equation}
a=\frac{M_{1}M_{2}}{2},b=-1,c=M_{1}M_{2},z=\frac{-2M_{1}}{M_{1}+M_{2}}.
\end{equation}
One can then calculate the extended recurrence relation%
\begin{align}
&  16\left(  \frac{2M_{1}}{M_{1}+M_{2}}+1\right)  \left(  -\frac{M_{1}+M_{2}%
}{2}\right)  ^{2}\left(  \frac{\epsilon}{2}\cos\phi\right)  ^{2p_{1}}%
A_{1}\nonumber\\
&  =8\cdot2^{P_{1}}\left(  \frac{M_{1}M_{2}}{2}+2\right)  \left(  \frac
{2M_{1}}{M_{1}+M_{2}}+2\right)  \left(  -\frac{M_{1}+M_{2}}{2}\right)  \left(
\frac{\epsilon}{2}\cos\phi\right)  ^{p_{1}+1}A_{2}\nonumber\\
&  -2^{P_{1}}\left(  M_{1}M_{2}+2\right)  \left(  \frac{\epsilon}{2}\cos
\phi\right)  ^{2}A_{3}%
\end{align}
where $p_{1}$ is an arbitrary integer. More extended recurrence relations can
be similarly derived.

The existence of these low-energy stringy symmetries comes as a surprise in terms of the perspective of
Gross's high-energy symmetries \cite{GM,Gross,GrossManes}.
Finally, in contrast to the Regge string spacetime symmetry, which\ was shown
to be related to $SL(5,C)$ of the Appell function $F_{1}$, we found that
the low-energy stringy symmetry is related to $SL(4,C)$ \cite{sl4c} of the
Gauss hypergeometry functions $_{2}F_{1}.$

\subsection{Summary}

In this section, we rederive from the LSSAs the relations or symmetries among
SSAs of different string states at three different scattering limits. We first
reproduce the linear relations \cite{CHLTY2,CHLTY1} of the HSSA from the LSSA
in the hard scattering limit. We also obtain Appell functions $F_{1}$ and
Gauss hypergeometric functions $_{2}F_{1}$ with $SL(5,C)$ and $SL(4,C)$
symmetry in the Regge and the nonrelativistic limits, respectively. In contrast
to the linear relations in the hard scattering limit, we obtain extended recurrence relations for the cases of RSSAs and NSSAs. These
two classes of recurrence relations are closely related to those of the LSSAs
with $K=2$ and $K=1$, respectively. In the end, we also show that with the
nonrelativistic string BCJ relations \cite{LLY}, the extended recurrence
relations we obtained can be used to connect SSAs with different spin states
and different channels.

\section{Conclusions and Future Works}

In this review, we provide a different perspective to demonstrate the Gross conjecture regarding the
high-energy symmetry of string theory \cite{GM,GM1,Gross,Gross1,GrossManes}.
We review our recent construction of the exact SSAs of three tachyons and one
arbitrary string state, or the LSSAs, in the $26D$ open bosonic string theory.
In addition, we discover that these LSSAs form an infinite-dimensional
representation of the $SL(K+3,%
\mathbb{C}
)$ group. Moreover, we show that the $SL(K+3,%
\mathbb{C}
)$ group can be used to solve all the LSSAs and express them in terms of one amplitude.

As an important application in the hard scattering limit, the LSSAs can be used
to prove the Gross conjecture regarding the high-energy symmetry of string theory, which was
previously corrected and proved by the method of decoupling of zero norm
states (ZNSs)
\cite{Lee,LeePRL,lee-Ov,ChungLee1,ChanLee1,ChanLee,ChanLee2,CHL,CHLTY2,
CHLTY1,susy}. In this sense, the results of the LSSAs presented in this review
extend the Gross conjecture to all kinematic regimes. Finally, the exact LSSA can
be used to rederive the recurrence relations of SSAs in the Regge scattering
limit with associated $SL(5,%
\mathbb{C}
)$ symmetry and the extended recurrence relations (including the mass and spin
dependent string BCJ relations) in the nonrelativistic scattering limit with
associated $SL(4,%
\mathbb{C}
)$ symmetry. These results were first discovered without knowing the exact LSSA.

There are many important related issues that remain to be studied. To name
some examples, how can the LSSA be generalized to multitensor cases? Can one
calculate exactly five-point, six-point and even higher point functions for
arbitrary higher spin string states? Solving these issues would be important to
uncover the whole spacetime symmetry structure of string theory. Presumably,
the $SL(K+3,%
\mathbb{C}
)$ symmetry of the LSSA is only a small part of the whole spacetime symmetry
of string theory.

Another important issue is the construction of massive fermion SSAs for the
R-sector of superstrings. Recently, the present authors calculated a class of
polarized fermion string scattering amplitudes (PFSSAs) at arbitrary mass
levels \cite{LLYNEW}. They discovered that, in the hard scattering limit, the
functional forms of the non-vanishing PFSSAs at each fixed mass level are
independent of the choices of spin polarizations. This result agrees with
the Gross conjecture regarding the high-energy string scattering amplitudes extended to the
R-sector. In addition, this peculiar property of hard PFSSAs should be compared
with the usual spin polarization-dependence of the hard-polarized fermion
field theory scatterings. However, the construction of the PFSSA involved only
the leading Regge trajectory fermion string state of the R sector
\cite{Osch,RRR}. It is a nontrivial task to construct the general massive
fermion string vertex operators \cite{m1,m2,m3,m4}.

Many questions related to the construction of SSA involving the general
massive fermion string states need to be answered before we can better
understand the high-energy behavior of superstring theory.

\acknowledgments{We would like to thank H. Kawai, Y. Kentaroh, Taejin. Lee, Y. Okawa, T. Okuda
and C.I. Tan for discussions which helped to clarify many issues of the LSSA.
This work is supported in part by the Ministry of Science and Technology
(MoST) and S.T. Yau center of National Chiao Tung University (NCTU), Taiwan.}

\appendix

\setcounter{equation}{0} \renewcommand{\theequation}{\Alph{section}\arabic{equation}}

\section{Lauricella String Scattering Amplitudes}

In this appendix, we give a detailed calculation of the LSSA presented in the
text. We begin with a simple case of the four-point function with the three
tachyons and the highest spin state at mass level $M_{2}^{2}=2(N-1)$,
$N=p+q+r$ with the following form:
\begin{equation}
\left\vert p,q,r\right\rangle =\left(  \alpha_{-1}^{T}\right)  ^{p}\left(
\alpha_{-1}^{P}\right)  ^{q}\left(  \alpha_{-1}^{L}\right)  ^{r}|0,k\rangle.
\end{equation}
The $\left(  s,t\right)  $ channel of this scattering amplitude can be
calculated to be%
\begin{align}
A_{st}^{(p,q,r)}  &  =\frac{\sin(\pi k_{2}\cdot k_{4})}{\sin(\pi k_{1}\cdot
k_{2})}A_{tu}^{(p,q,r)}=\frac{\sin(\frac{u}{2}+2-N)\pi}{\sin(\frac{s}%
{2}+2-N)\pi}A_{tu}^{(p,q,r)}\nonumber\\
&  =\frac{(-1)^{N}\Gamma(\frac{s}{2}+2-N)\Gamma(\frac{-s}{2}-1+N)}%
{\Gamma(\frac{u}{2}+2)\Gamma(\frac{-u}{2}-1)}A_{tu}^{(p,q,r)}\nonumber\\
&  =\frac{(-1)^{N}\Gamma(\frac{s}{2}+2-N)\Gamma(\frac{-s}{2}-1+N)}%
{\Gamma(\frac{u}{2}+2)\Gamma(\frac{-u}{2}-1)}\nonumber\\
&  \times\int_{1}^{\infty}dx\,x^{k_{1}\cdot k_{2}}(x-1)^{k_{2}\cdot k_{3}%
}\cdot\left[  \frac{k_{1}^{T}}{x}+\frac{k_{3}^{T}}{x-1}\right]  ^{p}%
\nonumber\\
&  \cdot\left[  \frac{k_{1}^{P}}{x}+\frac{k_{3}^{P}}{x-1}\right]  ^{q}%
\cdot\left[  \frac{k_{1}^{L}}{x}+\frac{k_{3}^{L}}{x-1}\right]  ^{r}\nonumber\\
&  =\frac{\Gamma(\frac{s}{2}+2-N)\Gamma(\frac{-s}{2}-1+N)}{\Gamma(\frac{u}%
{2}+2)\Gamma(\frac{-u}{2}-1)}\left(  -k_{3}^{T}\right)  ^{p}\left(  -k_{3}%
^{P}\right)  ^{q}\left(  -k_{3}^{L}\right)  ^{r}\nonumber\\
&  \times\int_{1}^{\infty}dx\,x^{k_{1}\cdot k_{2}}(x-1)^{k_{2}\cdot k_{3}%
}\cdot\left(  1-(\frac{-k_{1}^{T}}{k_{3}^{T}}))\frac{x-1}{x}\right)
^{p}\nonumber\\
&  \cdot\left(  1-(\frac{-k_{1}^{P}}{k_{3}^{P}})\frac{x-1}{x}\right)
^{q}\cdot\left(  1-(\frac{-k_{1}^{L}}{k_{3}^{L}})\frac{x-1}{x}\right)  ^{r}.
\end{align}
In the above calculation, we have used the string BCJ relation:
\cite{stringBCJ,stringBCJ2,LLY}%
\begin{equation}
A_{st}^{(p,q,r)}=\frac{\sin(\pi k_{2}\cdot k_{4})}{\sin(\pi k_{1}\cdot k_{2}%
)}A_{tu}^{(p,q,r)}.
\end{equation}
The next step is to perform a change of variable $\frac{x-1}{x}=x^{\prime}$ to get%

\begin{align}
A_{st}^{(p,q,r)}  &  =\frac{\Gamma(\frac{s}{2}+2-N)\Gamma(\frac{-s}{2}%
-1+N)}{\Gamma(\frac{u}{2}+2)\Gamma(\frac{-u}{2}-1)}\left(  -k_{3}^{T}\right)
^{p}\left(  -k_{3}^{P}\right)  ^{q}\left(  -k_{3}^{L}\right)  ^{r}\nonumber\\
&  \times\int_{0}^{1}dx^{\prime}\,x^{\prime\frac{-t}{2}-2}(1-x^{\prime
})^{\frac{-u}{2}-2}\left(  1-(\frac{-k_{1}^{T}}{k_{3}^{T}})x^{\prime}\right)
^{p}\nonumber\\
&  \cdot\left(  1-(\frac{-k_{1}^{P}}{k_{3}^{P}})x^{\prime}\right)  ^{q}%
\cdot\left(  1-(\frac{-k_{1}^{L}}{k_{3}^{L}})x^{\prime}\right)  ^{r}%
\nonumber\\
&  =\frac{\Gamma(\frac{s}{2}+2-N)\Gamma(\frac{-s}{2}-1+N)}{\Gamma(\frac{u}%
{2}+2)\Gamma(\frac{-u}{2}-1)}\nonumber\\
&  \cdot\left(  -k_{3}^{T}\right)  ^{p}\left(  -k_{3}^{P}\right)  ^{q}\left(
-k_{3}^{L}\right)  ^{r}\frac{\Gamma(\frac{-t}{2}-1)\Gamma(\frac{-u}{2}%
-1)}{\Gamma(\frac{s}{2}+2-N)}\nonumber\\
&  \times F_{D}^{(3)}(\frac{-t}{2}-1,-p,-q,-r,\frac{s}{2}+2-N;\frac{-k_{1}%
^{T}}{k_{3}^{T}},\frac{-k_{1}^{P}}{k_{3}^{P}},\frac{-k_{1}^{L}}{k_{3}^{L}}),
\end{align}
which can be written as%

\begin{align}
A_{st}^{(p,q,r)}  &  =\left(  -k_{3}^{T}\right)  ^{p}\left(  -k_{3}%
^{P}\right)  ^{q}\left(  -k_{3}^{L}\right)  ^{r}\frac{\Gamma(\frac{-s}%
{2}-1+N)\Gamma(\frac{-t}{2}-1)}{\Gamma(\frac{u}{2}+2)}\nonumber\\
&  \times F_{D}^{(3)}(\frac{-t}{2}-1,-p,-q,-r,\frac{s}{2}+2-N;-C^{T}%
,-C^{P},-C^{L}) \label{111}%
\end{align}
if we define%
\begin{equation}
k_{i}^{X}=e^{X}\cdot k_{i}\text{, \ \ \ \ }\frac{k_{3}^{X}}{k_{1}^{X}}%
=C^{X}\text{.}%
\end{equation}

We are now ready to calculate the LSSA; namely, the string scattering
amplitude with three tachyons and one general higher spin state in
Equation (\ref{state}). The detailed calculation is as follows:
\begin{align}
A_{st}^{(p_{n};q_{m};r_{l})}  &  =\frac{\sin(\pi k_{2}\cdot k_{4})}{\sin(\pi
k_{1}\cdot k_{2})}A_{tu}^{(p_{n};q_{m};r_{l})}=\frac{\sin(\frac{u}{2}+2-N)\pi
}{\sin(\frac{s}{2}+2-N)\pi}A_{tu}^{(p_{n};q_{m};r_{l})}\nonumber\\
&  =\frac{(-1)^{N}\Gamma(\frac{s}{2}+2-N)\Gamma(\frac{-s}{2}-1+N)}%
{\Gamma(\frac{u}{2}+2)\Gamma(\frac{-u}{2}-1)}\nonumber\\
&  \cdot\int_{1}^{\infty}dx\,x^{k_{1}\cdot k_{2}}(1-x)^{k_{2}\cdot k_{3}}%
\cdot\prod_{n=1}\left[  \frac{\left(  -1\right)  ^{n-1}(n-1)!k_{1}^{T}}{x^{n}%
}+\frac{(-1)^{n-1}(n-1)!k_{3}^{T}}{(x-1)^{n}}\right]  ^{p_{n}}\nonumber\\
&  \cdot\prod_{m=1}\left[  \frac{\left(  -1\right)  ^{m-1}(m-1)!k_{1}^{P}%
}{x^{m}}+\frac{(-1)^{m-1}(m-1)!k_{3}^{P}}{(x-1)^{m}}\right]  ^{q_{m}%
}\nonumber\\
&  \cdot\prod_{l=1}\left[  \frac{\left(  -1\right)  ^{l-1}(l-1)!k_{1}^{L}%
}{x^{l}}+\frac{(-1)^{l-1}(l-1)!k_{3}^{L}}{(x-1)^{l}}\right]  ^{r_{l}%
}\nonumber\\
&  =\frac{(-1)^{N}\Gamma(\frac{s}{2}+2-N)\Gamma(\frac{-s}{2}-1+N)}%
{\Gamma(\frac{u}{2}+2)\Gamma(\frac{-u}{2}-1)}\nonumber\\
&  \int_{1}^{\infty}dx\,x^{k_{1}\cdot k_{2}}(1-x)^{k_{2}\cdot k_{3}-N}%
\cdot\prod_{n=1}\left(  k_{3}^{T}\left(  -1\right)  ^{n-1}(n-1)![1-(\frac
{-k_{1}^{T}}{k_{3}^{T}})(\frac{x-1}{x})^{n}]\right)  ^{p_{n}}\nonumber\\
&  \cdot\prod_{m=1}\left(  k_{3}^{P}\left(  -1\right)  ^{m-1}(m-1)![1-(\frac
{-k_{1}^{P}}{k_{3}^{P}})(\frac{x-1}{x})^{m}]\right)  ^{q_{m}}\nonumber\\
&  \cdot\prod_{l=1}\left(  k_{3}^{L}\left(  -1\right)  ^{l-1}(l-1)![1-(\frac
{-k_{1}^{L}}{k_{3}^{L}})(\frac{x-1}{x})^{l}]\right)  ^{r_{l}}.
\end{align}
We can then perform a change of variable $\frac{x-1}{x}=y$ to get
\begin{align}
A_{st}^{(p_{n};q_{m};r_{l})}  &  =\frac{(-1)^{N}\Gamma(\frac{s}{2}%
+2-N)\Gamma(\frac{-s}{2}-1+N)}{\Gamma(\frac{u}{2}+2)\Gamma(\frac{-u}{2}%
-1)}\int_{0}^{1}dy\,y^{k_{2}\cdot k_{3}-N}(1-y)^{-k_{1}\cdot k_{2}-k_{2}\cdot
k_{3}+N-2}\nonumber\\
&  \cdot\prod_{n=1}\left(  k_{3}^{T}\left(  -1\right)  ^{n-1}(n-1)![1-(\frac
{-k_{1}^{T}}{k_{3}^{T}})y^{n}]\right)  ^{p_{n}}\nonumber\\
&  \cdot\prod_{m=1}\left(  k_{3}^{P}\left(  -1\right)  ^{m-1}(m-1)![1-(\frac
{-k_{1}^{P}}{k_{3}^{P}})y^{m}]\right)  ^{q_{m}}\nonumber\\
&  \cdot\prod_{l=1}\left(  k_{3}^{L}\left(  -1\right)  ^{l-1}(l-1)![1-(\frac
{-k_{1}^{L}}{k_{3}^{L}})y^{l}]\right)  ^{r_{l}}\nonumber\\
&  =\frac{(-1)^{N}\Gamma(\frac{s}{2}+2-N)\Gamma(\frac{-s}{2}-1+N)}%
{\Gamma(\frac{u}{2}+2)\Gamma(\frac{-u}{2}-1)}\cdot\prod_{n=1}\left[  \left(
-1\right)  ^{n-1}(n-1)!k_{3}^{T}\right]  ^{p_{n}}\nonumber\\
&  \prod_{m=1}\left[  \left(  -1\right)  ^{m-1}(m-1)!k_{3}^{P}\right]
^{q_{m}}\prod_{l=1}\left[  \left(  -1\right)  ^{l-1}(l-1)!k_{3}^{L}\right]
^{r_{l}}\nonumber\\
&  \cdot\int_{0}^{1}dy\,y^{k_{2}\cdot k_{3}-N}(1-y)^{-k_{1}\cdot k_{2}%
-k_{2}\cdot k_{3}+N-2}\nonumber\\
&  \cdot\left(  1-(z_{n}^{T}y)^{n}\right)  ^{p_{n}}\left(  1-(z_{m}^{P}%
y)^{m}\right)  ^{q_{m}}\left(  1-(z_{l}^{L}y)^{l}\right)  ^{r_{l}}.
\end{align}
Finally the LSSA can be written in the following form:

\begin{align}
A_{st}^{(p_{n};q_{m};r_{l})}  &  =\frac{\Gamma(\frac{s}{2}+2-N)\Gamma
(\frac{-s}{2}-1+N)}{\Gamma(\frac{u}{2}+2)\Gamma(\frac{-u}{2}-1)}\prod
_{n=1}\left[  -(n-1)!k_{3}^{T}\right]  ^{p_{n}}\nonumber\\
&  \cdot\prod_{m=1}\left[  -(m-1)!k_{3}^{P}\right]  ^{q_{m}}\prod_{l=1}\left[
-(l-1)!k_{3}^{L}\right]  ^{r_{l}}\nonumber\\
&  \cdot\int_{0}^{1}dy\,y^{\frac{-t}{2}-2}(1-y)^{\frac{-u}{2}-2}[(1-z_{n}%
^{T}y)(1-z_{n}^{T}\omega_{n}{}^{2}y)...(1-z_{n}^{T}\omega_{n}^{n-1}y)]^{p_{n}%
}\nonumber\\
&  \cdot\lbrack(1-z_{m}^{P}y)(1-z_{m}^{P}\omega_{m}y)...(1-z_{m}^{P}\omega
_{m}^{m-1}y)]^{q_{m}}\nonumber\\
&  \cdot\lbrack(1-z_{l}^{L}y)(1-z_{l}^{L}\omega_{l}y)...(1-w_{l}^{L}\omega
_{l}^{l-1}y)]^{p_{n}},
\end{align}
which can then be written in terms of the D-type Lauricella function
$F_{D}^{(K)}$ as follows:
\begin{align}
&  A_{st}^{(p_{n};q_{m};r_{l})}\nonumber\\
&  =\frac{\Gamma(\frac{s}{2}+2-N)\Gamma(\frac{-s}{2}-1+N)}{\Gamma(\frac{u}%
{2}+2)\Gamma(\frac{-u}{2}-1)}\frac{\Gamma(\frac{-t}{2}-1)\Gamma(\frac{-u}%
{2}-1)}{\Gamma(\frac{s}{2}+2-N)}\nonumber\\
&  \cdot\prod_{n=1}\left[  -(n-1)!k_{3}^{T}\right]  ^{p_{n}}\prod_{m=1}\left[
-(m-1)!k_{3}^{P}\right]  ^{q_{m}}\prod_{l=1}\left[  -(l-1)!k_{3}^{L}\right]
^{r_{l}}\nonumber\\
&  \cdot F_{D}^{(K)}\left(
\begin{array}
[c]{c}%
-\frac{t}{2}-1;\left\{  -p_{1}\right\}  ^{1},...,\left\{  -p_{n}\right\}
^{n},\left\{  -q_{1}\right\}  ^{1},...,\left\{  -q_{m}\right\}  ^{m},\left\{
-r_{1}\right\}  ^{1},...,\left\{  -r_{l}\right\}  ^{l};\frac{s}{2}+2-N;\\
\left[  z_{1}^{T}\right]  ,...,\left[  z_{n}^{T}\right]  ,\left[  z_{1}%
^{P}\right]  ,...,\left[  z_{m}^{P}\right]  ,\left[  z_{1}^{L}\right]
,...,\left[  z_{l}^{L}\right]  ,
\end{array}
\right) \nonumber\\
&  =\frac{\Gamma(\frac{-s}{2}-1+N)\Gamma(\frac{-t}{2}-1)}{\Gamma(\frac{u}%
{2}+2)}\prod_{n=1}\left[  -(n-1)!k_{3}^{T}\right]  ^{p_{n}}\prod_{m=1}\left[
-(m-1)!k_{3}^{P}\right]  ^{q_{m}}\prod_{l=1}\left[  -(l-1)!k_{3}^{L}\right]
^{r_{l}}\nonumber\\
&  \cdot F_{D}^{(K)}\left(
\begin{array}
[c]{c}%
-\frac{t}{2}-1;\left\{  -p_{1}\right\}  ^{1},...,\left\{  -p_{n}\right\}
^{n},\left\{  -q_{1}\right\}  ^{1},...,\left\{  -q_{m}\right\}  ^{m},\left\{
-r_{1}\right\}  ^{1},...,\left\{  -r_{l}\right\}  ^{l};\frac{s}{2}+2-N;\\
\left[  z_{1}^{T}\right]  ,...,\left[  z_{n}^{T}\right]  ,\left[  z_{1}%
^{P}\right]  ,...,\left[  z_{m}^{P}\right]  ,\left[  z_{1}^{L}\right]
,...,\left[  z_{l}^{L}\right]
\end{array}
\right)  . \label{2222}%
\end{align}
In the above calculation, we have defined%
\begin{equation}
k_{i}^{X}\equiv e^{X}\cdot k_{i}\text{, }\omega_{k}=e^{\frac{2\pi i}{k}%
}\text{\ , \ \ }z_{k}^{X}=(\frac{-k_{1}^{X}}{k_{3}^{X}})^{\frac{1}{k}}%
\end{equation}
and%
\begin{equation}
\left\{  a\right\}  ^{n}=\underset{n}{\underbrace{a,a,\cdots,a}}\text{,
\ \ }\left[  z_{k}^{X}\right]  =z_{k}^{X},z_{k}^{X}e^{\frac{2\pi i}{k}}%
,\cdots,z_{k}^{X}e^{\frac{2\pi i(k-1)}{k}}\text{ or }z_{k}^{X},z_{k}^{X}%
\omega_{k},...,z_{k}^{X}\omega_{k}^{k-1}.
\end{equation}
The integer $K$ in Equation (\ref{2222}) is defined to be%
\begin{equation}
\text{ }K=\underset{\{\text{for all }r_{j}^{T}\neq0\}}{\sum j}%
+\underset{\{\text{for all }r_{j}^{P}\neq0\}}{\sum j}+\underset{\{\text{for
all }r_{j}^{L}\neq0\}}{\sum j}.
\end{equation}
For a given $K$, there can be an LSSA with a different mass level $N$.

Alternatively, by using the identity of the Lauricella function for $b_{i}\in
Z^{-}$%
\begin{align}
&  F_{D}^{(K)}\left(  a;b_{1},...,b_{K};c;x_{1},...,x_{K}\right)
=\frac{\Gamma\left(  c\right)  \Gamma\left(  c-a-\sum b_{i}\right)  }%
{\Gamma\left(  c-a\right)  \Gamma\left(  c-\sum b_{i}\right)  }\nonumber\\
\cdot &  F_{D}^{(K)}\left(  a;b_{1},...,b_{K};1+a+\sum b_{i}-c;1-x_{1}%
,...,1-x_{K}\right)  ,
\end{align}
we can rederive the string BCJ relation \cite{stringBCJ, stringBCJ2,LLY}%
\begin{align}
\frac{A_{st}^{(r_{n}^{T},r_{m}^{P},r_{l}^{L})}}{A_{tu}^{(r_{n}^{T},r_{m}%
^{P},r_{l}^{L})}}  &  =\frac{(-)^{N}\Gamma\left(  -\frac{s}{2}-1\right)
\Gamma\left(  \frac{s}{2}+2\right)  }{\Gamma\left(  \frac{u}{2}+2-N\right)
\Gamma\left(  -\frac{u}{2}-1+N\right)  }\nonumber\\
&  =\frac{\sin\left(  \frac{\pi u}{2}\right)  }{\sin\left(  \frac{\pi s}%
{2}\right)  }=\frac{\sin\left(  \pi k_{2}\cdot k_{4}\right)  }{\sin\left(  \pi
k_{1}\cdot k_{2}\right)  }. \label{BCJ5}%
\end{align}
Equation (\ref{BCJ5}) gives another form of the $\left(  s,t\right)  $ channel
amplitude,
\begin{align}
&  A_{st}^{(r_{n}^{T},r_{m}^{P},r_{l}^{L})}\nonumber\\
&  =B\left(  -\frac{t}{2}-1,-\frac{s}{2}-1\right)  \prod_{n=1}\left[
-(n-1)!k_{3}^{T}\right]  ^{r_{n}^{T}}\nonumber\\
&  \cdot\prod_{m=1}\left[  -(m-1)!k_{3}^{P}\right]  ^{r_{m}^{P}}\prod
_{l=1}\left[  -(l-1)!k_{3}^{L}\right]  ^{r_{l}^{L}}\nonumber\\
&  \cdot F_{D}^{(K)}\left(  -\frac{t}{2}-1;R_{n}^{T},R_{m}^{P},R_{l}^{L}%
;\frac{u}{2}+2-N;\tilde{Z}_{n}^{T},\tilde{Z}_{m}^{P},\tilde{Z}_{l}^{L}\right)
\label{st}%
\end{align}
and similarly the $\left(  t,u\right)  $ channel amplitude
\begin{align}
&  A_{tu}^{(r_{n}^{T},r_{m}^{P},r_{l}^{L})}\nonumber\\
&  =B\left(  -\frac{t}{2}-1,-\frac{u}{2}-1\right)  \prod_{n=1}\left[
-(n-1)!k_{3}^{T}\right]  ^{r_{n}^{T}}\nonumber\\
&  \cdot\prod_{m=1}\left[  -(m-1)!k_{3}^{P}\right]  ^{r_{m}^{P}}\prod
_{l=1}\left[  -(l-1)!k_{3}^{L}\right]  ^{r_{l}^{L}}\nonumber\\
&  \cdot F_{D}^{(K)}\left(  -\frac{t}{2}-1;R_{n}^{T},R_{m}^{P},R_{l}^{L}%
;\frac{s}{2}+2-N;Z_{n}^{T},Z_{m}^{P},Z_{l}^{L}\right)  . \label{tu5}%
\end{align}

In Equations (\ref{st}) and (\ref{tu5}), we have defined%
\begin{equation}
R_{k}^{X}\equiv\left\{  -r_{1}^{X}\right\}  ^{1},\cdots,\left\{  -r_{k}%
^{X}\right\}  ^{k}\text{ with }\left\{  a\right\}  ^{n}%
=\underset{n}{\underbrace{a,a,\cdots,a}},
\end{equation}
and%
\begin{equation}
Z_{k}^{X}\equiv\left[  z_{1}^{X}\right]  ,\cdots,\left[  z_{k}^{X}\right]
\text{ with }\left[  z_{k}^{X}\right]  =z_{k0}^{X},\cdots,z_{k\left(
k-1\right)  }^{X}%
\end{equation}
where%
\begin{equation}
z_{k}^{X}=\left\vert \left(  -\frac{k_{1}^{X}}{k_{3}^{X}}\right)  ^{\frac
{1}{k}}\right\vert ,\ z_{kk^{\prime}}^{X}=z_{k}^{X}e^{\frac{2\pi ik^{\prime}%
}{k}},\ \tilde{z}_{kk^{\prime}}^{X}\equiv1-z_{kk^{\prime}}^{X}%
\end{equation}
for $k^{\prime}=0,\cdots, k-1.$

Finally, by using the notation introduced above, the $(s,t)$ channel amplitude
in\linebreak Equation (\ref{2222}) can then be rewritten as%
\begin{align}
&  A_{st}^{(r_{n}^{T},r_{m}^{P},r_{l}^{L})}\nonumber\\
&  =B\left(  -\frac{t}{2}-1,-\frac{s}{2}-1+N\right)  \prod_{n=1}\left[
-(n-1)!k_{3}^{T}\right]  ^{r_{n}^{T}}\nonumber\\
&  \cdot\prod_{m=1}\left[  -(m-1)!k_{3}^{P}\right]  ^{r_{m}^{P}}\prod
_{l=1}\left[  -(l-1)!k_{3}^{L}\right]  ^{r_{l}^{L}}\nonumber\\
&  \cdot F_{D}^{(K)}\left(  -\frac{t}{2}-1;R_{n}^{T},R_{m}^{P},R_{l}^{L}%
;\frac{s}{2}+2-N;Z_{n}^{T},Z_{m}^{P},Z_{l}^{L}\right)  . \label{tu}%
\end{align}


\end{document}